\documentclass[a4paper,12pt, epsfig]{article}
\usepackage{epsfig}
\usepackage{amssymb}
\usepackage{amsfonts}

\newskip\humongous \humongous=0pt plus 1000pt minus 1000pt

\newif\ifdtup

\catcode`\@=11

\@addtoreset{equation}{section}
\def\theequation{\thesection.\arabic{equation}}

\def\@normalsize{\@setsize\normalsize{15pt}\xiipt\@xiipt
\abovedisplayskip 14pt plus3pt minus3pt%
\belowdisplayskip \abovedisplayskip
\abovedisplayshortskip \z@ plus3pt%
\belowdisplayshortskip 7pt plus3.5pt minus0pt}

\def\small{\@setsize\small{13.6pt}\xipt\@xipt
\abovedisplayskip 13pt plus3pt minus3pt%
\belowdisplayskip \abovedisplayskip
\abovedisplayshortskip \z@ plus3pt%
\belowdisplayshortskip 7pt plus3.5pt minus0pt
\def\@listi{\parsep 4.5pt plus 2pt minus 1pt
     \itemsep \parsep
     \topsep 9pt plus 3pt minus 3pt}}

\relax

\catcode`@=12

\catcode`\@=11

\def\section{\@startsection{section}{1}{\z@}{3.5ex plus 1ex minus
   .2ex}{2.3ex plus .2ex}{\large\bf}}

\def\thesection{\arabic{section}}
\def\thesubsection{\arabic{section}.\arabic{subsection}}

\def\appendix{\setcounter{section}{0}
 \def\thesection{Appendix \Alph{section}}
 \def\thesubsection{\Alph{section}.\arabic{subsection}}
 \def\theequation{\Alph{section}.\arabic{equation}}}

\def\YGrule{0.4}   
\def\YGbox{6.5}    
\def\SymBoxes#1#2#3#4{\newdimen\un@t \un@t#3%
\raisebox{#1}{\rule{#2\un@t}{#4}\hskip-#2\un@t
\@tempdimb\un@t \advance\@tempdimb by-#4\@tempcntb#2\relax%
\@whilenum{\@tempcntb>0}\do{
\rule{#4}{\un@t}\hskip\@tempdimb \advance\@tempcntb by\m@ne}%
\hskip-#2\un@t \rule[\un@t]{#2\un@t}{#4}%
\rule[\un@t]{#4}{#4}\hskip-#4
\rule{#4}{\un@t}}\hskip-#4}                
\def\Young{\@ifnextchar[{\@Young}{\@Young[0]}}
\def\@Young[#1]#2{\newdimen\YG@unit \YG@unit\YGbox pt%
\newdimen\h@ight \h@ight#1\YG@unit \@tempcnta-1\relax
\@tfor\c@ount:=#2\do{\advance\@tempcnta by\@ne}
\@tempdima\@tempcnta\YG@unit%
\advance\h@ight by\@tempdima\relax     
\@tfor\c@ount:=#2\do{\SymBoxes{\h@ight}{\c@ount}{\YG@unit}{\YGrule pt}%
\@tempdima-\c@ount\YG@unit \hskip\@tempdima%
\advance \h@ight by -\YG@unit}         
\@tempdima\YG@unit \multiply\@tempdima by\@car#2\@nil %
\hskip\@tempdima}                      
\def\YoungTab{\@ifnextchar[{\@YoungIdx}{\@YoungIdx[0]}}
\def\@YoungIdx[#1]{\@ifnextchar[{\@iYoungIdx[#1]}{\@iYoungIdx[#1][\@empty]}}
\def\@iYoungIdx[#1][#2]#3{%
\newdimen\YG@unit \YG@unit\YGbox pt\newdimen\YG@rule \YG@rule \YGrule pt
\newcount\c@ount \c@ount\z@ \newdimen\skip@wd \unitlength\@ne pt
\newdimen\h@ight \h@ight#1\YG@unit \@tempcnta\m@ne\relax
\@tfor\d@um:=#3\do{\advance\@tempcnta by\@ne}
\@tempdima\@tempcnta\YG@unit%
\advance\h@ight by\@tempdima\relax
\@tfor\@idxlist:=#3\do{
\@tempcnta\z@\hskip.5\YG@rule\relax
\@for\@idx:=\@idxlist\do{
\raisebox{\h@ight}{\makebox(\YGbox,\YGbox){#2$\@idx$}}
\advance\@tempcnta by\@ne}\hskip-.5\YG@rule%
\@tempdima-\@tempcnta\YG@unit \hskip\@tempdima%
\ifnum\c@ount=\z@ \skip@wd-\@tempdima\fi \relax
\SymBoxes{\h@ight}{\@tempcnta}{\YG@unit}{\YG@rule}%
\hskip\@tempdima \advance\h@ight by -\YG@unit
\advance\c@ount by\@ne}
\hskip\skip@wd}                      

\begin{document}

\newcommand{\beq}{\begin{equation}}
\newcommand{\eeq}{\end{equation}}
\newcommand{\bea}{\begin{eqnarray}}
\newcommand{\eea}{\end{eqnarray}}
\newcommand{\beas}{\begin{eqnarray*}}
\newcommand{\eeas}{\end{eqnarray*}}
\newcommand{\defi}{\stackrel{\rm def}{=}}
\newcommand{\non}{\nonumber}
\newcommand{\bquo}{\begin{quote}}
\newcommand{\enqu}{\end{quote}}
\def\de{\partial}
\def\Tr{ \hbox{\rm Tr}}
\def\const{\hbox {\rm const.}}
\def\o{\over}
\def\im{\hbox{\rm Im}}
\def\re{\hbox{\rm Re}}
\def\bra{\langle}\def\ket{\rangle}
\def\Arg{\hbox {\rm Arg}}
\def\Re{\hbox {\rm Re}}
\def\Im{\hbox {\rm Im}}
\def\diag{\hbox{\rm diag}}
\def\longvert{{\rule[-2mm]{0.1mm}{7mm}}\,}
\def\a{\alpha}
\def\dag{{}^{\dagger}}
\def\tq{{\widetilde q}}
\def\p{{}^{\prime}}
\def\W{W}
\def\N{{\cal N}}
\newcommand{\Z}{\ensuremath{\mathbb Z}}
\begin{titlepage}
\begin{flushright}
ULB-TH/04-30\\
hep-th/0411074\\
\end{flushright}
\bigskip
\def\thefootnote{\fnsymbol{footnote}}

\begin{center}
{\large  {\bf
Monopoles Can be Confined by 0, 1 or 2 Vortices
 } }
\end{center}

\bigskip
\begin{center}
{\large  Roberto AUZZI \footnote{\texttt{ r.auzzi@sns.it}}$^{(1,2)}$ , Stefano BOLOGNESI\footnote{\texttt{ s.bolognesi@sns.it}} $^{(1,2)}$, \\
 Jarah EVSLIN\footnote{\texttt{ jevslin@ulb.ac.be}} $^{(3)}$
 \vskip 0.10cm
 }
\end{center}

\renewcommand{\thefootnote}{\arabic{footnote}}

\begin{center}
{\it   \footnotesize
Scuola Normale Superiore - Pisa,
 Piazza dei Cavalieri 7, Pisa, Italy $^{(1)}$ \\
\vskip 0.10cm
Istituto Nazionale di Fisica Nucleare -- Sezione di Pisa, \\
     Via Buonarroti, 2, Ed. C, 56127 Pisa,  Italy $^{(2)}$\\
\vskip 0.10cm
International Solvay Institutes,\\
Physique Th\'eorique et Math\'ematique,\\
Universit\'e Libre
de Bruxelles,\\C.P. 231, B-1050, Bruxelles, Belgium $^{(3)}$}

\end {center}

\noindent
\begin{center} {\bf Abstract} \end{center}
There are three types of monopole in gauge theories with fundamental matter and ${\cal N}=2$ supersymmetry broken by a superpotential.  There are unconfined 0-monopoles and also 1 and 2-monopoles confined respectively by one or two vortices transforming under distinct components of the unbroken gauge group.  If a Fayet-Iliopoulos term is added then there are only 2-monopoles.  Monopoles transform in the bifundamental representation of two components of the unbroken gauge symmetry, and if two monopoles share a component they may form a boundstate.  Selection rules for this process are found, for example vortex number is preserved modulo 2.  We find the tensions of the vortices, which are in general distinct, and also the conditions under which vortices are mutually BPS.  Results are derived in field theory and also in MQCD, and in quiver theories a T-dual picture may be used in which monopoles are classified by quiver diagrams with two colors of vertices.

\vfill

\end{titlepage}

\bigskip

\hfill{}
\bigskip

\section{Introduction }

Our ancestors once dreamed of understanding strong interactions by understanding the worldsheet theory of the vortices that confine quarks.  In ${\cal N}=2$ supersymmetric gauge theories with fundamental hypermultiplets this program has recently been reborn as the worldsheet theories have been identified as linear sigma models.  While the vortices have been constructed semiclassically, the linear sigma models encode information about the quantum theory, such as the BPS spectrum and Seiberg duality \cite{dorey,DHT,HT,SY-vortici,HT2}.
 Similar vortices are found also in supersymmetric theories in six spacetime dimensions
with fundamental hypermultiplets \cite{nitta1,nitta2}.

\begin{figure}[ht]
\begin{center}
\leavevmode
\epsfxsize 15  cm
\epsffile{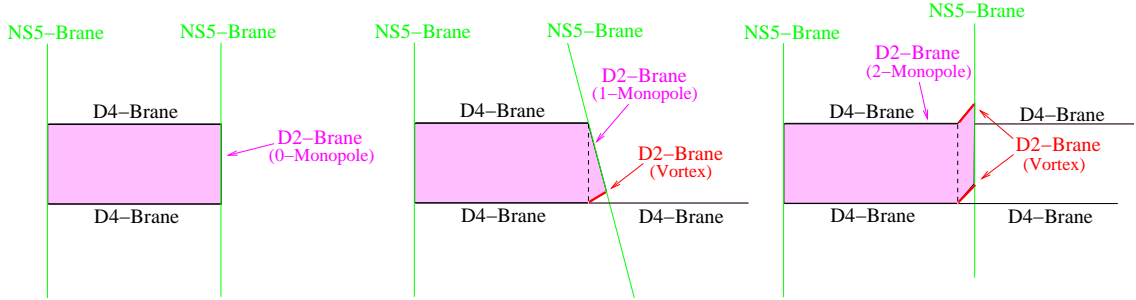}
\end{center}
\caption{\footnotesize Here are brane cartoons for the 0,1,2-monopoles in a $U(2)$ gauge theory.  The 't Hooft-Polyakov monopoles are D2-branes bounded by the D4 and NS5-branes, while the vortices are the continuations of these D2-branes that must exist if one of the D4-NS5 corners is ruptured, which corresponds to a superpotential or FI term.}
\label{branette}
\end{figure}

So far the spectrum of monopoles and vortices has only been found in a few special kinds of theories, such as those with an FI term \cite{HT,Shifman-yung1,HT2} and those where the ${\cal N}=2$ SUSY is softly broken to ${\cal N}=1$ by a superpotential that is quadratic in the adjoint chiral superfield \cite{Yung:2000uy,VY,MY,Shifman-yung2,vortici,monovortice}.  In both of these cases bare mass terms for the quarks may be used to break the gauge symmetry and thus create 't Hooft-Polyakov monopoles and their nonabelian generalizations \cite{GoddardNuytsOlive,Weinberg,monopoli}.  However the two cases yield different patterns of confinement.  In the first case, in which every color is locked to a flavor, monopoles are each confined by two vortices of equal tension, which are mutually BPS if they approach the 2-monopole from opposite directions \cite{tong-monopolo}.  In the second case, in which all of the bare masses are taken to be equal and so only one unbroken component of the color group may be locked to a flavor group, each monopole is confined by a single vortex, which is a vortex in the squark condensate of the locked component.  Although they can form stable, spinning mesons, these 1-monopoles can never be BPS, as they are pulled in a single direction by a single vortex.  However if there is a second unlocked, unbroken color component then the spectrum will also include unconfined abelian 't Hooft-Polyakov 0-monopoles that transform under the unlocked $U(1)\times U(1)$. We will see that the brane cartoons of Figure \ref{branette} realize these three possibilities. Here monopoles are excitations of D2-branes bounded by a pair of D4 and NS5-branes.  A nontrivial superpotential or FI term can rupture the boundary by separating a D4 and NS5, and in this case the D2-brane must continue along the rupture forming a strip in one of the gauge theory directions.  This strip yields the gauge theory vortex.

In this note we will see 
that in  $\mathcal{N}=2$ $U(N)$ gauge theories
softly broken by a superpotential that is a nondegenerate polynomial in the adjoint chiral multiplet the three types of monopoles coexist. Every species of monopole is constructed from two components of the unbroken gauge group corresponding to two distinct eigenvalues of the adjoint scalar $\phi$. In a supersymmetric vacuum every component of the unbroken gauge group must correspond to an eigenvalue of the adjoint scalar that is either semiclassically equal to an extremum of the superpotential or else to the bare mass of a sufficient number of quarks.  In the second case we say that the colors are locked to the flavors of these quarks.  In Fig.~\ref{branette} we see that locking occurs when a D4 continues past an NS5.  If both eigenvalues employed by a monopole are minima of the superpotential then the monopole is not confined.  However for each eigenvalue that is locked to a flavor there is a single vortex in the VEV of the corresponding squark that confines the monopole.  If a color eigenvalue matches a bare mass and is also an extremum of the superpotential then we say that the color is only marginally locked as it unlocks under a small perturbation of the superpotential.  In this case the corresponding squark has no VEV and so produces no vortices.

 In the case of $U(N)$ gauge theories we may generalize this situation by including an FI term $r\neq 0$, in which case all colors must be locked ($\phi_i=m_i$ where $m_i$
are the squark hypemultiplet masses) and all squarks have nontrivial VEVs. All monopoles are confined by two vortices, although the tensions of these two vortices may differ in the presence of a nontrivial superpotential.  
In general we find that semiclassically each vortex 
is approximately BPS and has tension:
\beq
T_i=4\pi\sqrt{|W^{\prime}(m_i)|^2+r^2} \ ,\label{ten}
\eeq
where $W$ is the superpotential.
 We will see that this is consistent with the $SU(2)$ $R$-symmetry at a fixed value of $\phi$. In fact it is proportional to the absolute value of the meson VEV of the condensate in which it is a vortex, which suggests the quantum generalization. 
 Eq.~(\ref{ten}) has a non-BPS classical correction which vanishes in the following weak-coupling limit:
\beq
e^2|{{W\p}\p}(m_i)^2 {W\p}(m_i)^2| << (|{W\p}(m_i)|^2+r^2)^{3/2}\ 
\eeq
where $e^2$ is the gauge coupling constant.
Notice that there are no classical corrections to the tension Eq.~(\ref{ten})
if $W''(m_i) = 0$ and $r,W'(m_i) \neq 0$
and also  if $W'(m_i)=0$ and $r,W''(m_i) \neq 0$.
 A conjectured quantum version of Eq.~(\ref{ten}) has appeared in Ref.~\cite{Stefano}.

In addition to a field theory demonstration of the properties of the vortices in $U(N)$ gauge theories, we provide an argument via an MQCD brane cartoon that allows us to easily extend our results to $SO(N)$ and $SP(N)$ theories and to understand the report between vortices and monopoles.  In the semiclassical regime of the gauge theory one may ignore $g_s$ corrections to the supergravity configuration and so one NS5-brane is flat and located at $x^7=x^8=x^9=0$ while the other is located at
\beq
x^7=r,\quad x^8+ix^9=W^\prime(x^4+ix^5)\ .
\eeq
The tension (\ref{ten}) is then just the Pythagorean theorem for the length of a D2-brane stretched along a line segment in the $x^{7-9}$ space.

As usual quiver gauge theories are created by compactifying the $x^6$ direction which means that flavor branes now have a finite length and so the flavor symmetry is gauged.  While unlocked colors are still D4-branes that stretch from one NS5 brane to the other, locked colors are D4-branes that wrap the entire circle $x^6$.  T-dualizing this circle we find an ALE space in which locked colors are D3-branes and unlocked colors and flavors are fractional D3-branes.  The superpotential and FI terms blow up the singularities and as usual the fractional D3-branes blow up into D5-branes wrapping the resolved 2-cycles.  The monopoles are D2-branes in the original IIA picture, and they T-dualize to 3-branes that wrap these 2-cycles and connect two color branes.  If the color brane at one end of a monopole is locked then it is a 3-brane and does not wrap the 2-cycle, and so the monopole, which does wrap the 2-cycle, cannot end on the color brane and must continue.  This continuation is the vortex.  The fact that the vortex wraps the 2-cycle whose area scales with the superpotential and FI term is consistent with Eq.~(\ref{ten}).  On the other hand if the 2-cycle degenerates at this point then the 3-brane can bound the monopole and so there is no vortex, as expected in the marginally locked case.

We find some new (not present when flavor is not gauged) bound states, such as the always unconfined 't Hooft-Polyakov 0-monopole of the locked color-flavor group, which is in general a boundstate of a 't Hooft-Polyakov 2-monopole from the broken color group with one from the broken flavor group.  In the type IIB brane cartoon these correspond to D-strings which in most of moduli space are away from the singularity and so are described by an ${\mathcal N}=4$ theory in the infrared.  We may encode the monopole spectrum in a quiver diagram in which there are two colors of vortex, red for the locked colors and black for the unlocked colors.  The edges correspond to monopoles, and there is one vortex ending on a given monopole for each red vortex on which it ends.  Boundstates correspond to paths constructed from the edges.

Throughout this note we will restrict consideration to nondegenerate superpotentials, and so unlocked color symmetries are dynamically broken to their maximal torus.  Thus while $U(N), SO(N),$\ and $SP(N)$\ monopoles transform under the bifundamental representation of the two color components with which they are constructed, these components must each either be locked or else be single $U(1)$'s.  Only marginally locked components may yield unconfined nonabelian 0-monopoles, although without finetuning it is possible that quantum corrections will either break the nonabelian symmetry or lead to vortices with a tension of order the QCD scale $\Lambda$.  On the other hand in quiver gauge theories the above ${\mathcal N}=4$ monopoles are always unconfined and their potentially nonabelian symmetry is not dynamically broken.

In Section~\ref{campi} we define the crucial concept of color-flavor locking and then find vortices and their tensions directly in field theory.  The tension formula for vortices in $U(N)$ gauge theories is reproduced using Hanany-Witten brane cartoons in Section~\ref{mqcdsezione}.  This approach allows us to find the number of vortices that confine each monopole, selection rules for monopole interactions and a list of charges.  These results are then generalized to $SO(N)$ and $SP(N)$ gauge theories and to dyons.  Finally in Section~\ref{faretrasezione} we consider quiver gauge theories, in which a similar brane cartoon yields the same tension formula for vortices.  Quiver diagrams with two colors of vortices classify the monopoles in quiver gauge theories as well as encoding the number of vortices that confine each monopole.

\section{Nonabelian Vortices} \label{campi}

We begin by reviewing the semiclassical construction of vortices in several supersymmetric gauge theories.  These theories are not all asymptotically free, but they can be UV completed by adding locked pairs of colors and sufficiently massive flavors.  We may restrict our attention to theories that are free in the IR by locking nonabelian color groups to sufficient numbers of flavors, thus the semiclassical solutions (often semilocal vortices \cite{semi1,semi2,semi3}) may be expected to survive quantum corrections.  In addition these vortices survive quantum corrections because they are BPS saturated, although their tensions are corrected \cite{Stefano}.

We will consider ${\cal N}=2$ superQCD, which contains vectormultiplets and fundamental hypermultiplets.  We will then include an FI term (for $U(N)$ theories) and or a superpotential, which may break the supersymmetry to ${\cal N}=1$ and lead to the existence of vortices.  Each species of vortex is constructed using squarks with a bare mass equal to a particular eigenvalue of the adjoint scalar VEV, and so there will be at most one species for each eigenvalue.  In the semiclassical regime considered here each vortex is constructed using only the terms in the Lagrangian involving the subset of squarks and the superpartners of the adjoint scalar with the given eigenvalue.  In particular the superpotential may be expanded about the eigenvalue and we will find that semiclassically only its first derivative affects the solution.  The tension of the vortex is then determined by the FI term and the derivative of the superpotential at the eigenvalue, which together form a triplet of the $SU(2)$ R-symmetry as they are the scalars of an ${\cal N}=2$ linear multiplet.  Thus for any eigenvalue we may use the R-symmetry to transform away the superpotential and leave only an FI term, and so the construction of any individual vortex will reduce to the case with ${\cal N}=2$ supersymmetry and no superpotential.  Of course if we consider two different types of vortices at once, for example because we are interested in their interactions or in a 2-monopole that is confined by both, then we cannot necessarily simultaneously rotate away the superpotential evaluated at the eigenvalues corresponding to both vortices.  Such a simultaneous rotation exists precisely for pairs of vortices that would be mutually BPS were they parallel or antiparallel, such as those of Refs.~\cite{MY,tong-monopolo}.

\subsection{Color-flavor locking}

Consider  ${\cal N}=2$, $U(N_c)$ gauge theory with $N_f$ flavours, 
broken to ${\cal N}=1$ with a superpotential $W(\Phi)$ and an FI term.
We will say that a color and flavor are locked if, including quantum corrections, the corresponding eigenvalue $\phi$ of the adjoint scalar is equal to the bare mass of the hypermultiplet flavor.  If the FI term vanishes and $\phi$ is an extremum of the superpotential then we will say that the color and flavor are not locked but only marginally locked.  In this case a small perturbation to the superpotential that shifts the critical point allows the vacuum to perturb to one in which the color and flavor and truly locked.
However while the color and flavor appear to be locked, an arbitrarily small change in the superpotential or bare quark mass may also separate them.

 If the eigenvalues $\phi$ of the adjoint scalar and $m$ of the bare mass matrix have degeneracies $N_c$ and $N_f$ respectively then they may define a low energy $U(N_c)$ gauge theory with $N_f$ flavors of massless hypermultiplets.  If $N_f\geq 2N_c$ then this theory is IR free and in some vacua the $U(N_c)$ gauge symmetry is exact and the eigenvalues $\phi$ do not receive quantum corrections.  In these cases all $N_c$ colors are necessarily locked.  More generally there is dynamical symmetry breaking and only a subset of the eigenvalues are unaffected and correspondingly a subset of the colors whose eigenvalues are semiclassically degenerate is in fact locked \cite{APS,CKM}.

In the remainder of this section we will see that vortices can be constructed whenever a color $a$ and flavor $i$ are locked, but as the locking becomes marginal the vortex delocalizes out of existence because the limiting configuration is homogeneous.  Each fundamental vortex is constructed from a locked color-flavor pair, where the charge is the winding number of the corresponding squark VEV.  This VEV may be nonvanishing when the squark mass
\beq
m_{ai}=|\phi_a-m_i|\
\eeq
vanishes, that is when $\phi_a$ is equal to the bare mass $m_i$.  If there are extra flavors at the same mass in addition to color-flavor pairs then at some points on the Higgs branch non-BPS semi-local vortices may be constructed as well.  In this paper we will concern ourselves only with BPS local vortices.

All of the abelian and nonabelian 't Hooft-Polyakov monopoles that we will consider in the main body of this note are constructed from a $U(2)$ broken to a $U(1)^2$ gauge symmetry, and thus they are constructed from only two colors which may then be embedded into a larger symmetry group.  We will see in the next section that these monopoles are confined by a number of vortices equal to the number of these two colors that are locked.

\subsection{Abelian vortex with FI term} \label{fisubsec}

The $SU(2)$ R-symmetry may be used to rotate the superpotential at any single eigenvalue $\phi_a$ into an FI term, and the gauge symmetry may be used to rotate a nonabelian vortex (see \cite{vortici}) solution into a given $U(1)$ locked color and flavor.  Thus all of the vortices considered in this note may be constructed from the abelian vortex of ${\cal N}=2$ SQED with an FI term $2r$.  We will follow the construction in, for example, Ref.~\cite{HT}.

In superfield notation the Lagrangian density is
\bea
{\cal L}&=&\int d^2\theta \, \frac{1}{4e^2}W^{\a}W_{\a}+h.c.\\
&&+\int d^2\theta d^2 \bar{\theta} \,(\frac{1}{e^2}\Phi\dag e^V\Phi e^{-V}+Q\dag e^VQ+{\widetilde Q}\dag e^{-V}{\widetilde Q})\nonumber\\
&&+\int d^2\theta \, \sqrt{2}({\widetilde Q}\Phi Q-m{\widetilde Q}Q)+h.c.
-\int d^2\theta d^2\bar{\theta} \, 2rV\ .\nonumber
\eea
This yields a bosonic potential for the selectrons $q,\ \tq$ and for $\phi$, the scalar ${\cal N}=2$ superpartner of the photon
\beq
V=2|(\phi-m)q|^2+2|(\phi-m)\tq|^2+2e^2|\tq q|^2+\frac{e^2}{2}(|q|^2-|\tq|^2-2r)^2\ . \label{pot}
\eeq
In the nonabelian case, near the core of a magnetic monopole, at distances much smaller than the inverse FI term, we may ignore the $r$ term in Eq.~(\ref{pot}) and the squark VEVs vanish.  Monopoles are the resulting nontrivial configurations of $\phi$ and the gauge fields.

At the larger distance scales at which monopoles are confined by vortices we may not ignore the FI parameter $2r$, and so the squark VEV cannot vanish as a result of the last term of the potential (\ref{pot}).  Thus the first two terms of the potential ensure that the VEV of $\phi$ is precisely fixed at the bare quark mass $m$.  In addition the third term implies that only one of $q$ and $\tq$ can be nontrivial, by convention we will say that it is $q$ (so $r>0$).  This means that vortices are nontrivial configurations of $q$ and the photon and so it suffices to consider only the terms in the Lagrangian built solely from the selectrons and photon
\beq
\label{BPSabelian}
{\cal L}\supset -\frac{1}{4e^2}F_{\mu\nu}F^{\mu\nu} - |D_\mu q|^2-\frac{e^2}{2}(q\dag q-2r)^2\ .
\eeq

If we consider a static string extended along the $3$-direction then only $F_{12}$ is nonvanishing.  The tension of such a string may be found by completing the square of the Hamiltonian density and integrating over a cross section
\beq
T=\int d^2 x \, \frac{1}{2e^2}(F_{12}+e^2(q\dag q-2r))^2+|{\cal D}_1q+i{\cal D}_2q|^2+{2r}F_{12}\ . \label{tens}
\eeq
The last term is proportional to the Chern class of the $U(1)$ gauge bundle over a compactification of the cross-section, and so is discrete.  Thus we may minimize the first term separately.  The first two terms are positive definite and the vortex may only be BPS saturated if they vanish, which yields the Bogomolny equations \cite{Bogomolny,HT}:
\beq
\label{BPSequations}
F_{12}=-e^2(q\dag q-2r)\ ,\qquad {\cal D}_1q=-i{\cal D}_2q \ .
\eeq
The tension of a BPS string is then equal to the topological term in Eq.~(\ref{tens}).  In particular the minimal string corresponds to a gauge bundle with Chern class $2\pi$ and so the minimal tension is
\beq
T={2r}\int F_{12}=4\pi r\ .
\eeq

\subsection{Nonabelian vortex}

Now we consider the general case of ${\cal N}=2$, $U(N_c)$ gauge theory with $N_f$ flavours, broken to ${\cal N}=1$ with a superpotential $W(\Phi)$ and an FI term $2r$. The Lagrangian is
\bea
\label{QFT}
{\cal L}&=&\int d^2\theta \,\frac{1}{2g^2}\Tr_{N_c} \, (W^{\alpha}W_{\alpha}) +h.c. \\ &&+ \int d^2\theta d^2\bar{\theta}  \,\frac{2}{g^2}\Tr_{N_c} \, (\Phi\dag e^V\Phi e^{-V})+ \int d^2\theta d^2\bar{\theta} \,\sum_{i=1}^{N_f} (Q_i\dag e^{V}Q^i+\widetilde{Q}_i e^{-V}\widetilde{Q}^{\dagger i} )\nonumber\\
&& + \int d^2\theta \,\sum_{i=1}^{N_f} \sqrt{2}(\widetilde{Q}_i\Phi Q^i - m_i\widetilde{Q}_i Q^i )+\sqrt{2} \Tr_{N_c} \, \W(\Phi)  +h.c.\nonumber\\
&&-\int d^2\theta d^2\bar{\theta} \, 2r \Tr_{N_c} \, V \ . \nonumber
\eea
The F terms for the squarks contribute to the bosonic potentials
\bea
&&V_{F_q}=2\sum_{i=1}^{N_f}\sum_{a=1}^{N_c}|\widetilde{q}^{ai}(\phi_a-m_i)|^2\ ,\\
&& V_{F_{\widetilde{q}}}=2\sum_{i=1}^{N_f}\sum_{a=1}^{N_c}|(\phi_a-m_i)q^{ai}|^2\ \nonumber
\eea
where we have used the D-term equations $[\phi,\phi\dag]=0$ to simultaneously diagonalize $\phi$ and $\phi\dag$.  In particular one can have quark condensate $q^{ai}\neq 0$ only if $\phi_a$ is equal to some mass $m_i$, in other words if there is color-flavor locking\footnote{If there are more flavors with this mass than the degeneracy of the $\phi$ eigenvalue then we find semilocal vortices.  See
\cite{AV, HT, HT2} for details.}.

If $N_j$ colors and flavors are locked at the same eigenvalue then we can build a nonabelian BPS vortex that confines the $U(N_j)$ magnetic flux. The $F_{\phi}$-term and the $D$-term together yield the potential for the squark fields
\beq
V= g^2 \Tr_{N_j} \, (|q\tq +\W'|^2)+\frac{g^2}{4}\Tr_{N_j} \, ((qq\dag -\tq\dag \tq-2r{\bf 1}_{N_j})^2)\ ,
\eeq
where we have suppressed the gauge and flavor indices, which are summed.
This may be expressed in a $SU(2)_R$ invariant form using the doublet $q^{\alpha}=(q,\tq\dag)$:
\beq
V=\frac{g^2}{2}\Tr_{N_j} \Tr_2 \, ({q\dag}^{\alpha}q_{\beta}-\frac{1}{2}\delta^{\alpha}_{\beta}{q\dag}^{\gamma} q_{\gamma}-\xi_a(\sigma_a)^\alpha_\beta)^2\ , \eeq
\[ -\xi_1+i\xi_2=\W\p(m)\ ,\qquad \xi_3=r\ .
\]
An $SU(2)_R$ rotation brings the potential to a  form with a new FI term and no superpotential
\beq
V=\frac{g^2}{4}\Tr_{N_j}\, ((qq\dag -\tq\dag \tq-2{r\p}{\bf 1}_{N_j})^2)\ ,
\eeq
where $r\p$ is
\beq
r\p=\sqrt{\W\p(m)^2+{r^2}}\ . \label{newv}
\eeq

As we will see in \ref{NAV}, the nonabelian vortex has tension
\beq
T=4\pi\sqrt{{\W\p}(m)^2+r^2}\ . \label{ten2}
\eeq
One might have expected the tension to scale with the dimension of the gauge group, because the Lagrangian terms
\beq
\dots+ \int d^2\theta \,\sqrt{2} \Tr_{N_c}\, \W(\Phi)  +h.c. \quad \textup{\ and\ } \quad -\int d^2\theta d^2\bar{\theta} 2r \Tr_{N_c}\, V
\eeq
are summed over gauge indices.  However we will see in brane cartoons that the tension, which is the width of a strip of membrane, is independent of the number of colors and flavors.  The nonabelian vortices are classically embeddings of single abelian vortices.

\subsection{Flux matching}

As a consistency check we will compare the vortex flux seen above to the flux sourced by a 2-monopole.
First we consider the simplest case, $U(2)$ broken to $U(1)\times U(1)$ by
\beq
\phi=\left(\begin{array}{cc}
\phi_1&\\
&\phi_2 \\
\end{array}\right) \ .
\eeq
Unlike vortices, which are constructed from the gluons and squarks, monopoles are configurations built from the gluons and the adjoint scalar and so are described by the Lagrangian terms
\beq
{\cal L} \supset -\frac{1}{2g^2}\Tr\,(F_{\mu\nu}F^{\mu\nu})-(D_{\mu}\phi)\dag (D^{\mu}\phi)\ .
\eeq
To calculate the flux we need only the configuration of the monopole at infinity. 
In the direction of the unit vector $\hat{r}$ the adjoint scalar asymptotes to
\beq
\phi(\hat r) \sim
\left(\begin{array}{cc} \phi_1+\phi_2/2&\\&\phi_1+\phi_2/2\\ \end{array}\right)+(\phi_1-\phi_2) \vec{\sigma} \cdot \hat r/2  \ ,
\eeq
while the gluons asymptote to
\beq
\vec{A}(\hat r) \sim -\frac{\vec{\sigma}\wedge\hat r}{2r} \ ,
\eeq
where $\vec{\sigma}$ are the Pauli matrices.
The magnetic field is
\beq
B_i=\frac 12 \epsilon_{ijk}F_{jk}=\frac{\hat{r}_i (\vec{\sigma}\cdot\hat r)}{2r^2} \ .
\eeq
We use a gauge transformation to orient $\phi$ in the direction $\sigma_3$. The $\vec B$ field transforms like $\phi$, so in this gauge the flux of the monopole is $2\pi$ times the generator $\sigma_3$:
\beq
\textup{flux}= \int_{{\cal S}^2} d \vec s \cdot \frac{\hat r}{2r^2} \sigma_3=2\pi\sigma_3 \ .
\eeq

Now we come to the vortex fluxes. We normalize the generators of the unbroken $U(1)$'s to be respectively $\frac1{\sqrt{2}}\left(\begin{array}{cc} 1&\\&0\\ \end{array}\right)$ and $\frac 1{\sqrt{2}}\left(\begin{array}{cc} 0&\\&1\\ \end{array}\right)$.
We know from Ref.~\cite{monovortice} that for a $U(1)$ vortex the unit of flux is $2\pi$ when the quarks have charge $1$.  Using our normalization the quark in the fundamental has charge $1/\sqrt{2}$, so the unit of flux is $2\sqrt{2}\pi$. Flux matching for 2-monopoles then reduces to the equality
\beq
\Phi_{\textup{\tiny{monopole}}}=2\pi\,\sigma_3=2\pi \,  \left(\begin{array}{cc} 1&\\&0\\ \end{array}\right)-2\pi\,\left(\begin{array}{cc} 0&\\&1\\ \end{array}\right)\ =\Phi_{\textup{\tiny{vortices}}}\ .
\eeq
Thus the flux sourced by a 2-monopole is entirely confined within its vortices.  While this is clearly not true for 0-monopoles, in $U(N)$ gauge theories it also fails for 1-monopoles.  However all of the flux is contained by the vortices of some of their 1-monopole cousins in $SU(N)$ gauge theories.  This has been demonstrated in Ref.~\cite{monovortice} for the $SU(2)$ monopoles of \cite{vortici}.

The nonabelian monopole is an embedding of the ordinary monopole $U(2) \to U(1) \times U(1)$.
The procedure to build the nonabelian vortex is analogous: embed the $U(1)$ vortex in the group $U(N_j)$.  Then the nonabelian flux matching is a simple consequence of the abelian matching.

\subsection{Non-BPS corrections}

Away from the weak coupling limit the tension has a non-BPS classical contribution when the 
second derivative of the superpotential is different from zero.
This contribution, and the limit in which it is small, has been studied in many classes of examples in Refs.~\cite{HSZ,VY,Hou,ken-rob,Stefano,prog}.  Here we provide a slight generalization of these discussions to the models of this note.  Such corrections do not threaten the stability of our vortices, which is a consequence of the topologically nontrivial winding of the squark field on an asymptotic circle.


Consider first $\N=2$  SQED with supersymmetry broken to $\N=1$ by a generic superpotential and with no FI term. One may obtain the BPS Lagrangian (\ref{BPSabelian}) with $r=|W\p(m)|$ by dropping all but the linear and constant terms in the superpotential and making  an $SU(2)_R$ rotation, or equivalently by using the BPS approximation
\beq
\label{BPSansatz}
\phi=m\ , \qquad \tq = -q\dag   \frac{W\p}{|W\p|} \ .
\eeq
This approximation does not satisfy the equation of motion for $\phi$
\beq
\label{phieq}
\frac{\Box \phi}{e^2}=2(\phi-m)(|q|^2+|\tq|^2)+2e^2 {W\p}\p(\phi)\dag (\tq q+W\p(\phi))\ .
\eeq
In particular combining Eqs.~(\ref{BPSansatz}) and (\ref{phieq}) one would obtain
\beq
0=e^2{W\p}\p(m)\dag (|q|^2-|W\p(m)|)
\eeq
which is false when ${W\p}\p\neq 0$ because inside the core of a vortex  $|q|^2-|W\p(m)|$ is nonzero.

Without the approximation (\ref{BPSansatz}) the tension is
\bea
\label {nonBPS}
T & =& \int d^2 x \,  \frac{1}{4e^2}F_{kl}F_{kl}+\frac{1}{e^2}\de_k\phi\dag\de_k\phi+(D_{k}q)^{\dagger}(D_{k}q)+(D_{k}\tq)^{\dagger}(D_{k} \tq) \\
&&+ 2|(\phi-m)q|^2+2|(\phi-m)\tq|^2+2e^2|\tq q+W\p(\phi)|^2+\frac{e^2}{2}(|q|^2-|\tq|^2)^2 \ .\nonumber
\eea
 We want to compute the leading correction to the BPS tension and, comparing it to the BPS tension,  determine the regime in which  the non-BPS contribution can be neglected.  The stationary equations deriving from (\ref{nonBPS}) are:
\bea
\label{nonBPSequation}
&\triangle \phi/e^2=2(\phi-m)(|q|^2+|\tq|^2)+2e^2 {W\p}\p(\phi)\dag (\tq q+W\p(\phi))\ .& 
\eea
To compute the first order correction we use the following strategy. First we insert the equations of motion (\ref{BPSequations}) and  then we search for the leading correction to the solutions. The quantities $q,\tq$\ and $\de_kF_{kl}$ 
only have corrections at a higher order, 
thus the lowest order correction is given by $\phi=m+\phi_{(1)}$
and by the second term of Eq.~(\ref{nonBPSequation})
 (the first term gives a correction of higher order):
\beq
\label{firstcorrection}
\triangle \phi_{(1)} = 2e^4 {W\p}\p(m)\dag (\tq q+W\p(m)).
\eeq

From this equation we are able to estimate $\phi_{(1)}$. Outside the radius of the vortex $R_v$, $\phi_{(1)}$ is small, while inside  $\phi_{(1)} \sim e^4 {W\p}\p W\p {R_{v}}^2$ (remember that $R_v \sim 1/e\sqrt{|W\p|}$).
The first order  correction to the tension comes from three pieces: the first is the kinetic term of $\phi$
\beq
\label{onecor}
\int d^2 x \, \frac{1}{e^2} \de_k \phi_{(1)}\dag \de_k\phi_{(1)} \sim e^6 |{{W\p}\p}^2 {W\p}^2| {R_{v}}^4 \sim e^2 |{{W\p}\p}^2| \ ,
\eeq
 the second is the sum of the $F_q$ and $F_{\tq}$ terms
\beq
\label{twocor}
\int d^2x \,  2|\phi_{(1)}q|^2+2|\phi_{(1)}\tq|^2  \sim e^8 |{{W\p}\p}^2 {W\p}^3| {R_{v}}^6 \sim e^2 |{{W\p}\p}^2| \ 
\eeq
and the last  is the deformation of the $F_{\phi}$ term
{\small \beq
\label{threecor}
\int d^2x \,  2e^2 (|\tq q+W\p(m+\phi_{(1)})|^2-
|\tq q+W\p(m)|^2)
\sim e^6 |{{W\p}\p}^2 {W\p}^2| {R_{v}}^4 \sim e^2 |{{W\p}\p}^2| \ .
\eeq}
All three of these corrections are of the same order and so the holomorphic tension is a good approximation to the real tension if the following condition is satisfied:
\beq
\label{goodone}
e^2|{{W\p}\p}^2| << |{W\p}| \ .
\eeq
See \cite{Hou} for a numerical computation of the first order correction that we have estimated above.

Let us now include the effect of the Fayet-Iliopolous term $r$.
The first order correction to the adjoint field is
given by (\ref{firstcorrection}). The estimate 
  $\phi_{(1)} \sim e^4 {W\p}\p W\p {R_{v}}^2$ is still valid,
  but now  $R_v \sim \frac{1}{e( |W\p|^2+r^2)^{1/4}}$.
  The first order  correction to the tension still comes from three pieces: the first is again the kinetic term of $\phi$
\beq
\label{onecorbis}
\int d^2 x \, \frac{1}{e^2} \de_k \phi_{(1)}\dag \de_k\phi_{(1)} \sim e^6 |{{W\p}\p}^2 {W\p}^2| {R_{v}}^4 \sim \frac{e^2|{{W\p}\p}^2 {W\p}^2|}{(|{W\p}|^2+r^2)} \ ,
\eeq
the second is again the sum of the $F_q$ and $F_{\tq}$ terms
\beq
\label{twocorbis}
\int d^2x \,  2|\phi_{(1)}q|^2+2|\phi_{(1)}\tq|^2  \sim e^8 {|{W\p}\p}^2 {W\p}^2| \sqrt{|{W\p}|^2 + r^2}
 {R_{v}}^6 \sim \frac{e^2|{{W\p}\p}^2 {W\p}^2|}{(|{W\p}|^2+r^2)} \ 
\eeq
and the last  is again the deformation of the $F_{\phi}$ term
{\small \beq
\label{threecorbis}
\int d^2x \,  2e^2 (|\tq q+W\p(m+\phi_{(1)})|^2-
|\tq q+W\p(m)|^2) \sim  e^6 |{{W\p}\p}^2 {W\p}^2| {R_{v}}^4 \sim \frac{e^2|{{W\p}\p}^2 {W\p}^2|}{(|{W\p}|^2+r^2)}  .
\eeq}
The leading correction to the tension is of order
   $\frac{e^2|{{W\p}\p}^2 {W\p}^2|}{(|{W\p}|^2+r^2)}$. 
So the condition (\ref{goodone}) is generalized to:
\beq
\label{fgoodone}
\frac{e^2|{{W\p}\p}^2 {W\p}^2|}{(|{W\p}|^2+r^2)} << \sqrt{(|{W\p}|^2+r^2)}\ .
\eeq
These results can be directly applied
in the $\N=2$ $U(N_c)$ gauge theory vortices studied
in the previous sections.  
The BPS approximation can be in general reliable in the weak coupling 
regime where $m >> \Lambda$ and so $e << 1$.
Notice also that there are no classical corrections to the BPS tension 
if $W''(m) = 0$ and $r,W'(m) \neq 0$
and also  if $W'(m)=0$ and $r,W''(m) \neq 0$ (see \cite{prog} for the detailed
analysis of an example of this last situation).

\section{MQCD} \label{mqcdsezione}

In MQCD, which is a realization of these field theories on the worldvolume of a brane configuration, we can easily reproduce the semiclassical tension formula (\ref{ten2}) and, although we will refrain, even compute its quantum corrections using the quantum moduli space explored in Ref.~\cite{HT}.  A formula for these corrections will be proposed in Ref.~\cite{Stefano}.  We will use MQCD to find the conditions under which monopoles are confined and to learn how many vortices confine them.  In addition we may use MQCD to learn that various combinations of monopoles may combine to form boundstates which are other monopoles, and as an application we will find selection rules that restrict how many vortices these monopoles may have.  Of course if the binding energy is nontrivial then the original configuration is non-BPS and so the dynamics of this physical process are out of the reach of MQCD, whose connection to 4-dimensional gauge theories relies heavily on supersymmetric nonrenormalization theorems.  We will also use MQCD to determine which vortices are mutually BPS, and finally to extend our results to $SO$ and $SP$ gauge theories.

\subsection{The Hanany-Witten cartoon for $\mathbf{N=2}$ superQCD}

Consider a stack of D4-branes orthogonal to two NS5-branes in type IIA string theory.  We may pretend that the spacetime is flat Minkowski space by using the coordinate redefinition in Ref.~\cite{Gibbons9803203}.  The NS5-branes extend along the directions $x^{0-5}$ and the D4-branes along $x^{0-3,6}$.  $N_c$ of the D4-branes are suspended between the NS5-branes while $N_f$ are semi-infinite, extending from one NS5-brane to infinity.  All branes are placed at $x^{7-9}=0$, although we will allow the D4-branes to move along the $x^{4,5}$ plane and the NS5-branes to move in the $x^6$-direction.

According to Refs.~\cite{hanany-witten,witten-n=2} this Hanany-Witten type brane configuration, in a wisely-chosen limit, yields a low energy 4-dimensional gauge theory that decouples from NS5-brane and gravitational degrees of freedom.  In particular the gauge theory enjoys ${\cal N}=2$ supersymmetry, a possibly broken $SU(N_c)$ gauge symmetry and $N_f$ flavors of fundamental hypermultiplets transforming under a possibly broken $U(N_f)$ global flavor symmetry.  Each of the ${N_{c}}^2-1$ vector multiplets comes with a complex scalar whose VEVs form a matrix whose $N_c$ eigenvalues $\phi_i$ are the positions of the suspended D4 color branes in the $x^{4,5}$ plane.  The $N_f$ positions of the semi-infinite D4 flavor branes are the bare quark hypermultiplet masses $m_i$.

\begin{figure}[ht]
\begin{center}
\leavevmode
\epsfxsize 9   cm
\epsffile{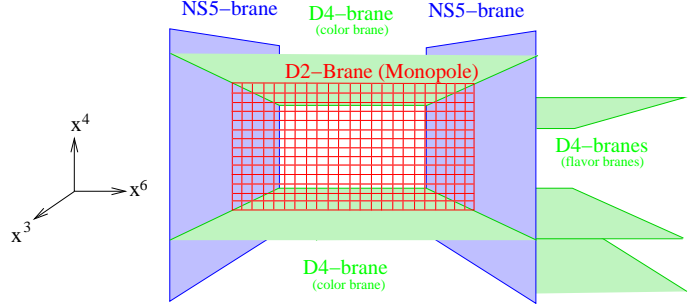}
\end{center}
\caption{ \footnotesize
A 't Hooft-Polyakov monopole in a theory with two Colors and three flavors.}
\label{monovort}
\end{figure}

\begin{figure}[ht]
\begin{center}
\leavevmode
\epsfxsize 10   cm
\epsffile{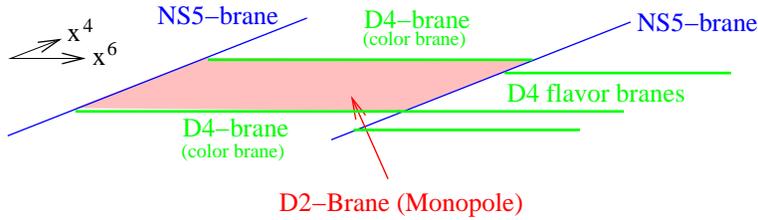}
\end{center}
\caption{ \footnotesize
This is another view of the previous figure.}
\label{monovort2}
\end{figure}

The gauge theory degrees of freedom are strings and D2-branes.  For example fundamental strings stretched between the color branes yield the adjoint vectormultiplets, while D2-branes that fill the rectangle between two color branes and the two NS5-branes are 't Hooft-Polyakov monopole hypermultiplets.  The quark hypermultiplets are not precisely fundamental strings as they lift to disks in M-theory and lie on top of an NS5-brane, where the string coupling is large and so perturbative string theory is invalid.  However in the semiclassical regime of large distances these strings approximate fundamental strings, for example their tensions asymptote to the string tension.

The semiclassical masses of the excitations are simply the areas of the corresponding membranes or lengths of the corresponding strings.  One may even incorporate quantum corrections by lifting this configuration to M-theory by turning on $g_s$, in which case the branes bend and the D2-brane areas change, realizing the instanton corrections to the semiclassical mass formulas.

\subsection{Confining monopoles}

\subsubsection*{ $U(2)$ gauge theory with an FI term}

\begin{figure}[ht]
\begin{center}
\leavevmode
\epsfxsize 10   cm
\epsffile{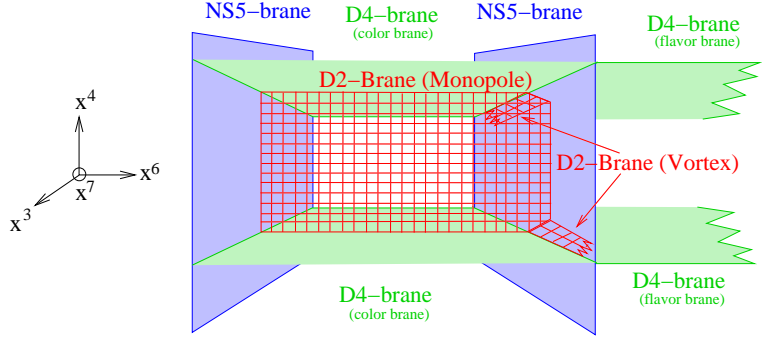}
\end{center}
\caption{\footnotesize
We have turned on an FI term, which corresponds to displacing the NS5 on the right.  The boundary of the monopole has now been broken in two places because the NS5 brane on the right no longer touches either of the color-flavor locked branes.  Each of these gaps in the D2-brane (the monopole) boundary is the beginning of a vortex, since it would be illegal (RR current conservation) for the 2-brane to end anywhere but on another brane.  This is the 1/4 BPS 2-monopole configuration of Ref.~\cite{tong-monopolo} in which the vortices leave the monopole in opposite directions.}
\label{FI}
\end{figure}

\begin{figure}[ht]
\begin{center}
\leavevmode
\epsfxsize 10   cm
\epsffile{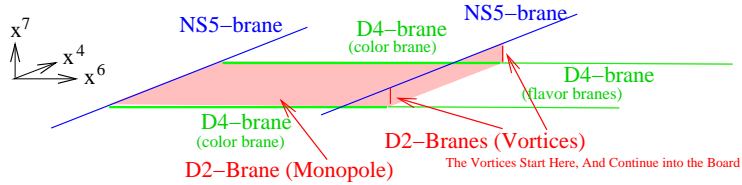}
\end{center}
\caption{ \footnotesize
This is Figure~\ref{FI} viewed from another perspective.}
\label{2dFI}
\end{figure}

The simplest example is a $U(2)$ gauge theory with an FI term.  The FI term means that the two NS5-branes, which are already separated in the $x^6$ direction, are now also relatively displaced in the $x^7$ direction.  If there are no flavors then we can rotate the coordinates so that the branes are again displaced in only the $x^6$ direction, although they are more distant than they were before and so the coupling has decreased.  This can be seen in field theory by writing the FI term as an integral over half of superspace and then completing the square by transforming the vector superfield and gauge coupling
\beq
W\longrightarrow W+\frac{r}{2}\ ,\qquad g^2\longrightarrow \frac{g^4}{g^2+r/2}\ ,
\eeq
and dropping the constant term.  Notice that this change of variables would not leave the Lagrangian invariant if one includes flavors of charged matter.

If we include flavors, by adding semi-infinite branes, then we have fixed a 6-direction and so no such rotation is possible.  Supersymmetry then requires that the color branes are orthogonal to $x^{7-9}$ and so if they begin on the leftmost NS5 then they do not intersect the rightmost NS5-brane.  The color branes must therefore continue forever on the right, or equivalently they are both locked to semi-infinite flavor branes.  Thus we need at least two flavors, corresponding to the extension of the two color branes past the NS5-brane that they miss.
The only 't Hooft-Polyakov monopole results from the broken $U(2)$.  It is bounded by the 2 color branes and 2 NS5-branes, and when the FI term vanishes this boundary is closed and the monopole is not confined.

When the FI term is nonzero the corners where the two D4-branes hit the displaced NS5-brane are no longer closed.  As a result the D2-branes continue past these corners, each in one of the spacetime directions.  More precisely RR charge conservation demands that D2-branes may only be bounded by other branes.  This is because $G_6$ is gauge-invariant and so $ddG_6=0$, but $ddF_6$ is a delta function on the boundary of a D2-brane and $F_6=G_6$ away from other branes, so the boundary is empty away from other branes.  To satisfy this we must include two strips which end on the rectangular parts of the monopole at the two broken corners, pasting the boundaries together so that the monopole no longer has any boundary off of the D4 and NS5-brane worldvolumes.  These strips extend along the 7-direction between the displaced NS5 and the corresponding D4's.  They each also extend along one of the space directions of the gauge theory and the time direction.  The thickness of both strips is the FI parameter, and so the monopoles are each confined by two vortices of equal tension.  This configuration, depicted\footnote{We thank David Tong for correcting a sign error in an earlier version of Figure~\ref{FI}.} in Figs.~\ref{FI} and \ref{2dFI}, is an example of a monopole of Ref.~\cite{HT}.

\subsubsection*{$U(2)$ gauge theory with adjoint chiral multiplet mass term}

\begin{figure}[ht]
\begin{center}
\leavevmode
\epsfxsize 10   cm
\epsffile{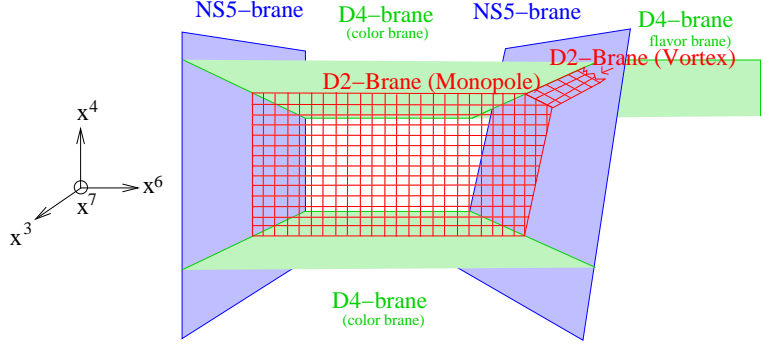}
\end{center}
\caption{\footnotesize
An adjoint chiral multiplet mass term has been turned on.  This corresponds to a rotation of the NS5-brane on the right.  As a result the monopole (a D2-brane) boundary is broken at a single corner.  D2-brane boundaries must lie along other branes, so the D2-brane continues away from the monopole.  This is the vortex.}
\label{mu}
\end{figure}

\begin{figure}[ht]
\begin{center}
\leavevmode
\epsfxsize 10   cm
\epsffile{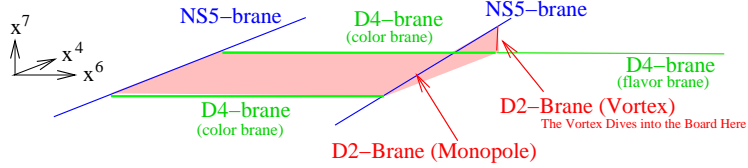}
\end{center}
\caption{ \footnotesize This is another view of Figure~\ref{mu}.}
\label{2dmu}
\end{figure}

Instead of an FI term we may include a mass $\mu$ for the adjoint chiral multiplet as in Figures~\ref{mu} and \ref{2dmu}.  Then both NS5-branes are at $x^7=0$ but one is at an angle
\beq
x^8+ix^9\sim\mu(x^4+ix^5)\ .
\eeq
A color brane extending along the $x^6$ direction may be suspended between the two NS5-branes when their $x^{8,9}$ coordinates agree, which means at $x^4=x^5=0$ corresponding to a vanishing VEV $\phi$.  However semiclassically a D4-brane corresponding to any nonvanishing eigenvalue $\phi$ will need to be locked to a flavor.  We may construct a monopole between any pair of D4-branes, and so we will consider a D2-brane that stretches between a D4-brane at $x^4=x^5=0$ and a D4-brane at some other point $x^4+ix^5=p$, where $p$ is the bare mass of the locked flavor.  The boundary of this D2-brane is again a rectangle when $\mu=0$, but when $\mu\neq 0$ only one of the corners breaks, the one at the intersection of $p$ and the rotated NS5.  Thus the conservation of RR charge implies that the D2-brane must continue past the broken corner, extending into a strip of vortex.  The width of this strip is $p\mu$, the distance between the rightmost NS5 and the locked D4-brane at its two ends, and so each monopole is confined by a single vortex with tension proportional to $p\mu$.  If $\Phi$ has another nonzero eigenvalue then there is another color-flavor locked pair and this theory will also contain a 2-monopole.

\subsubsection*{The general case}

We may now perturb the gauge theory by including a superpotential $\W$ and FI term $2r$.  In MQCD this is achieved by distorting one of the NS5-branes \cite{witten-n=1,Hori:1997ab,deBoer:1997ap}, for concreteness we will choose the rightmost ($x^6>0$) NS5-brane and we will let all of the flavor branes be semi-infinite on its right $(x^6\rightarrow +\infty)$.  Instead of being localized at the point $x^{7-9}=0$, the right NS5-brane now wraps
\beq
x^7=r,\qquad x^8+ix^9=\W\p(x^4+ix^5)\ . \label{ns5pos}
\eeq
This is consistent with the correspondence between rigid rotations of the coordinates $x^{7-9}$ and the broken $SU(2)$ R-symmetry of the gauge theory.
Supersymmetry requires that the D4-branes do not bend in the $x^{7-9}$ directions, and so at a value of $x^{4,5}$ such that the rightmost NS5 is at $x^{7-9}(x^4,x^5)\neq 0$, any supersymmetric D4-brane can not extend between the NS5-branes and so must be semi-infinite.  In particular a color brane suspended between the NS5-branes before the perturbation can no longer be suspended and so must connect to a flavor brane at the same $x^{4,5}$ coordinate, yielding color-flavor locking.

While color-flavor locking rescues the supersymmetry of the D4-branes, the boundary of a 't Hooft-Polyakov monopole D2-brane ending on such a D4-brane is broken.  In the unperturbed theory a corner of this D2-brane was the intersection of the D4 and the NS5.  However if the D4 and NS5 no longer intersect, because of an FI term or a superpotential whose derivative does not vanish at the corresponding eigenvalue $\phi$, the D2-brane which previously ended at this intersection no longer exists.  However a similar configuration does exist, in which the D2-brane still is bounded by the same two D4-branes and two NS5-branes, but at each broken corner a strip of D2-brane, extended between the D4 and NS5, continues into one of the $x^{1-3}$ directions of the worldvolume gauge theory.

This strip of D2-brane is the gauge theory vortex.  Its tension is the width of the strip, which is the distance between the D4-brane at $x^{7-9}=0$ and the NS5-brane whose position is given in Eq.~(\ref{ns5pos}).  This distance is
\beq
d=\sqrt{\Delta (x^7)^2+\Delta (x^8)^2+\Delta (x^9)^2}=\sqrt{|\W\p(x^4+ix^5)|^2+r^2}\ .
\eeq
Recalling that $x^4+ix^5$ is the eigenvalue of $\phi$ which is equal to the bare mass $m$ of the flavor to which it is locked, we see that this distance is, as desired, proportional to the vortex tension (\ref{ten2}).

Each monopole is constructed from a broken $U(2)$ color symmetry and thus ends on two D4-branes.  Thus each monopole potentially has two vortices, corresponding to the two corners at which the D4-branes intersect the rightmost NS5-brane.  However if a corner is unbroken the BPS mass of the vortex vanishes and the vortex becomes infinitely delocalized until only a constant configuration remains, that is to say that there is no vortex in the marginally locked case.  If there is an FI term then the two NS5-branes are always at different $x^7$ coordinates, thus both corners are broken and there are always two vortices.  If there is no FI term then monopoles are confined by a number of vortices equal to the number of eigenvalues of the $U(2)$ adjoint scalar at which $\W\p(\phi)$ is nonvanishing.

One might wonder what an FI term is in an $SU(N)$ gauge theory.  FI terms in such cartoons are D terms of a $U(1)$ extension on the $SU(N)$ gauge symmetry to $U(N)$.  In one UV completion, in which there are no more unlocked color branes or orientifold planes far away, the kinetic term of this $U(1)$ vectormultiplet is infinite \cite{witten-n=2}, corresponding to the fact that moving the center of mass of the color branes creates an infinite-energy distortion on the NS5-branes, and so the effective gauge symmetry is only $SU(N)$.  However the details of the relevant part of the MQCD brane cartoon are independent of the far away UV completion, and so the MQCD description does not distinguish between the case in which there is or is not a far away unlocked color that restores the full $U(N)$ symmetry.  This is not to say that the physics of the vortices is independent of the existence of this extra $U(1)$.  For example we will propose in \ref{suapp} that if all $N$ colors of an $SU(N)$ theory are locked, and so cannot contribute a dynamical $U(1)$, then two-monopoles cannot exist, and instead the flux may be entirely confined by one-monopoles.  Instead, as the existence of an FI parameter already suggests, we claim that MQCD teaches us about vortices in $U(N)$ gauge theories, and not in the $SU(N)$ gauge theories that we may have expected.  

In general monopoles may be charged under a residual nonabelian symmetry if some of the adjoint scalar VEVs $\phi$ are degenerate, or equivalently if the corresponding D4 color branes are coincident.  If this degeneracy persists in the quantum theory and if $W^{''}(\phi)\neq 0$ then supersymmetry requires that each of these colors locks to a flavor.  Again an FI term or $\W\p$ breaks a corner and the D2-brane must have a tendril that extends past this corner until it finds an antimonopole.  Thus the nonabelian monopoles of Refs.~\cite{GoddardNuytsOlive,Weinberg,vortici,HT} appear automatically and are in general confined.  This approach has still not been extended to different types of nonabelian monopoles in less supersymmetric theories, such as those of Refs.~\cite{Kneipp, Kneipp2,Kneipp3,Kneipp4,Kneipp5,AchucaUrre}.

\subsection{Mutually BPS vortices}

In addition to the tensions of the vortices, the brane cartoon also allows one to learn which vortices are mutually BPS.  A static vortex extends along the time direction, a gauge theory direction in $x^{1-3}$ and an internal direction in $x^{7-9}$ which is parallel to the $x^{7-9}$ displacement of the right NS5-brane.  Two vortices are only mutually BPS if these two direction vectors agree.  The sign of the gauge theory direction depends on the charge of the vortex, and the orientability of our D2-branes implies that the two vortices connected to a single monopole have opposite orientations.

For example consider the configuration of Refs.~\cite{HT,tong-monopolo} in which there is an FI term and no superpotential.  The FI term implies that all monopoles are confined by two vortices, each of which extends along the $x^7$ direction.  The opposite orientations of these two vortices then imply that they are mutually BPS only if they extend in opposite spatial directions $x^{1-3}$.  In this case not only are the vortices themselves mutually BPS, preserving two supercharges, but even the interface with the monopole preserves a supercharge.  In general this is not true, for example in the case of an $SU(2)$ gauge theory with an adjoint scalar mass and both color eigenvalues locked to flavors with bare masses $m$ and $-m$ we find two vortices with opposite orientations in the $x^{7-9}$ directions.  These vortices are mutually BPS if they extend from the monopole in the same spacetime direction, so that they have opposite spacetime $(x^{0-3})$ orientations.  However in this case they exert a net force on the monopole and so the monopole-vortex junction does not preserve any supersymmetry.  Conversely in Ref.~\cite{MY} collinear vortices with the same internal orientation are considered and it is found that they are mutually BPS when they have the same spacetime orientation, that is when their charges have the same sign.

\subsubsection*{Examples}

Different types of monopoles may appear, and interact, in the same theory.  For example consider a $U(4)$ gauge theory with the cubic superpotential
\beq
\W=\frac{\Phi^3}{3}-\Phi
\eeq
and two flavors of hypermultiplets with bare masses
\beq
m_1=2\ ,\qquad m_2=-2\ .
\eeq
This corresponds to a brane cartoon in which one NS5 brane is flat and located at $x^{7-9}$ while the other is at
\beq
x^8+ix^9=(x^4+ix^5)^2-1\ . \label{ufcurve}
\eeq
4 color branes are suspended between them and there are two semi-infinite branes extending from the rightmost (largest $x^6$) NS5-brane, the curved one, off to infinity on the right.

\begin{figure}[ht]
\begin{center}
\leavevmode
\epsfxsize 9   cm
\epsffile{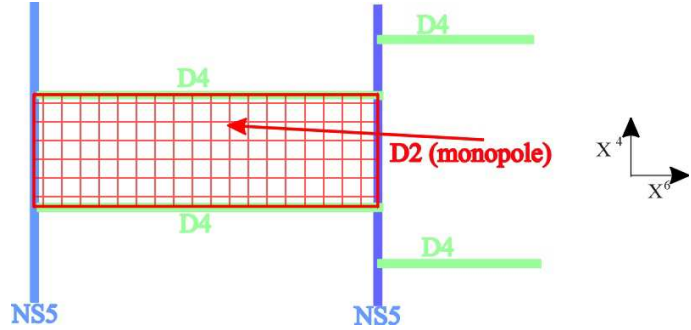}
\end{center}
\caption{\footnotesize Two color branes are suspended at $x^4= 1,\ x^5=0$ and two
are suspended at $x^4=-1,\ x^5=0$. There are classically four 0-monopoles of mass
2; another massless 0-monopole lives between the two color branes at each critical point but cannot be seen here as it has an expected area of zero.}
\label{esemp1}
\end{figure}

The two NS5-branes are at the same $x^{7-9}$ position, zero, at the zeros of Eq.~(\ref{ufcurve})
\beq
x^4=\pm 1\ ,\qquad x^5=0\ .
\eeq
Therefore in one possible vacuum the four color branes are suspended between the NS5-branes at these two points
(see Figure \ref{esemp1}).  If two are suspended at each critical point, this leads to a semiclassical $U(2)\times U(2)$ symmetry which is dynamically broken to $U(1)^4$ by quantum effects of order the QCD scale $\Lambda$.  Systematically ignoring the $1/g^2$ coefficients we find four unconfined 0-monopoles, all of mass $2+O(\Lambda)$, that span these two sets of color branes while a massless 0-monopole lives between the two color branes at each critical point.

\begin{figure}[ht]
\begin{center}
\leavevmode
\epsfxsize 9   cm
\epsffile{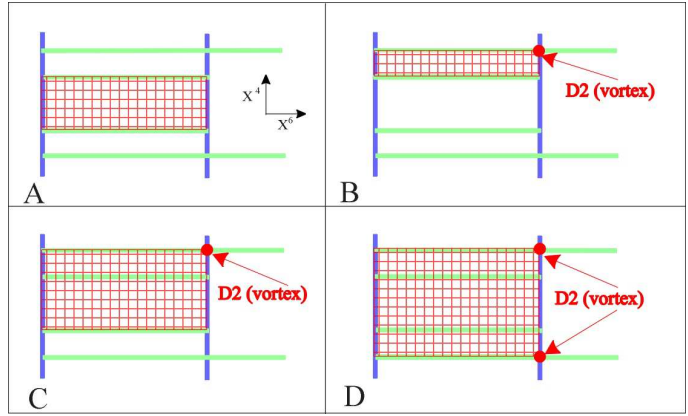}
\end{center}
\caption{\footnotesize
Color branes are suspended at $x^4=-2,-1,+1,+2,\ x^5=0$.
There is an unconfined mass 2 zero-monopole (A), there are two mass 1 (B) and two mass 3 one-monopoles (C) confined by a single vortex  and finally there is a mass 4 two-monopole (D).}
\label{esemp2}
\end{figure}

If we want to see vortices we need to lock some of the color branes to the flavor branes, which are at $x^4=\pm 2$.  For example, suppose that we lock each flavor brane to a color brane and put one of the remaining two color branes at each critical point, so that the color branes are all at $x^5=0$ but the $x^4$ coordinates of the four branes are $x^4=\pm 2$ and $x^4=\pm 1$.  Now we have a $U(1)^4$ gauge symmetry where the adjoint scalar has the four eigenvalues $\pm 2$ and $\pm 1$.  't Hooft-Polyakov monopoles exist between each pair of color branes.  There is an unconfined mass 2 zero-monopole, there are two mass 1 and two mass 3 one-monopoles confined by a single vortex of tension\footnote{We systematically ignore the factor of $4\pi$ in the vortex tension formula.} $\W\p(\pm 2)=3$ and finally there is a mass 4 two-monopole, whose two vortices each have a tension of 3 (see Figure \ref{esemp2}).

The monopoles are all mutually BPS, but some are marginally unstable and may decay into others.  For example the mass 4 two-monopole may decay into a mass 1 and a mass 3 one-monopole.  Note that the two original vortices are the two final vortices, and so an observer far away who only observes the vortices will not notice a difference.  In general such processes may also involve the creation or annihiliate of pairs of vortices and antivortices that are made of the same flavor of squark condensate.  The mass 3 one-monopole in turn may emit a mass 2 unconfined zero-monopole, leaving a confined mass 1 one-monopole attached to the original vortex.


\begin{figure}[ht]
\begin{center}
\leavevmode
\epsfxsize 9   cm
\epsffile{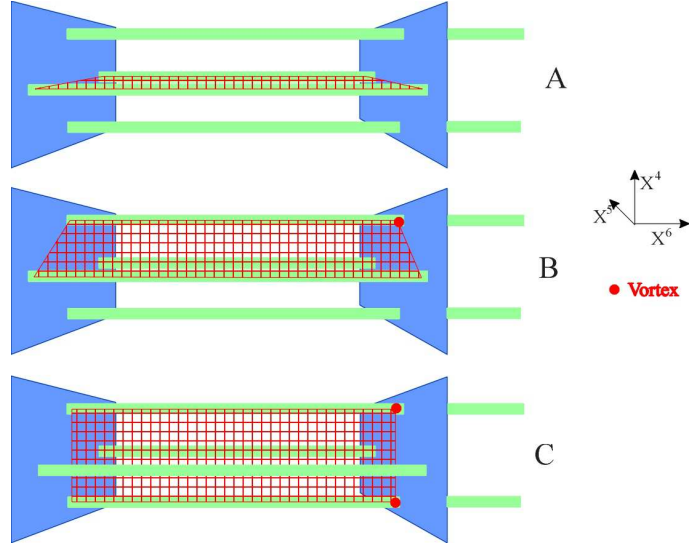}
\end{center}
\caption{\footnotesize
Color branes are suspended at $x^4=-2,+2,\quad x^5=0$ and at $x^4=0,\ x^5=-1,+1$.
There is an unconfined mass $2$ zero-monopole (A), there are four mass $\sqrt{5}$ one-monopoles (B) and there is a mass $4$ two-monopole (C). Some pairs of monopoles are not mutually BPS.}
\label{esemp3}
\end{figure}

In general the various monopoles need not be mutually BPS.  For example if we change the above superpotential to
\beq
\W=\frac{\Phi^3}{3}+\Phi
\eeq
then the NS5-branes will be adjacent at $x^4=0$, $x^5=\pm 1$.  We may consider the same configuration in which one of the four color branes goes to each critical point and one attaches to each flavor brane, but now the monopoles are no longer all parallel, and thus not mutually BPS (see Figure \ref{esemp3}).
 The masses of the 0 and 2-monopoles are unchanged, but the 1-monopoles now both have a  mass of $\sqrt{5}$ as the two types of color branes are further apart.  In addition the vortex tension has increased to 5, but this will not be relevant.  The key difference is that while two mass $\sqrt{5}$ monopoles may still combine to form a mass 2 or 4 monopole, the final state has a lower total mass and so there is now a binding energy.  In particular if the mass $\sqrt{5}$ monopoles do not share any $U(1)$, so that the vortices are different species, then they cannot combine\footnote{If they have the same orientation then they are mutually BPS.  If they have opposite orientations then they attract, but the lack of supersymmetry leaves the dynamics of their interaction beyond the reach of the MQCD approach employed here.  Perhaps a tachyon condensation based approach \cite{Sen} would be more applicable.}.  If they share a locked $U(1)$ then their vortices are of the same type with the opposite charge (we have chosen this convention for monopoles and antimonopoles) and so the two vortices annihilate, leaving the 0-monopole.  Finally if they share an unlocked $U(1)$ then the two vortices are of different types and the bound state is the 2-monopole, which has both of the vortices of the original monopoles.  If they had shared both $U(1)$'s then the two monopoles would be of the same type, and so they would combine marginally to form a charge 2 monopole.

\subsubsection*{Bound states}

From here the general pattern is easy to see.  The following combinations of monopoles may form a single monopole bound state if they share a single adjoint scalar eigenvalue.  Two $0$-monopoles, which are unconfined, may only combine to form another $0$-monopole.  A $0$-monopole and a $1$-monopole may form a $1$-monopole.  A $0$-monopole and a $2$-monopole cannot share any colors, as both colors of the first are unlocked and of the second are locked, and so there is no boundstate consisting solely of such a pair.  Two $1$-monopoles may combine to form a $0$-monopole or a $2$-monopole, depending on whether their vortices are of the same type.  A $1$-monopole and a $2$-monopole may combine to form another $1$-monopole.  Finally two $2$-monopoles may combine to form another $2$-monopole.

\begin{table}[h]
\begin{center}
\begin{tabular}{|l|l|l|l|}
\hline & \textbf{ 0-monopole} & \textbf{ 1-monopole} & \textbf{ 2-monopole} \\
\hline \textbf{  0-monopole }& 0-monopole & 1-monopole & no boundstate \\
\hline \textbf{ 1-monopole} & 1-monopole & 0 or 2-monopole & 1-monopole\\
\hline \textbf{ 2-monopole} & no boundstate & 1-monopole & 2-monopole \\
\hline
\end{tabular}
\end{center}
\caption{\footnotesize  Different combinations of monopoles may form a
single monopole bound state if they share a single adjoint scalar eigenvalue. }
\label{bbbb}
\end{table}

While we have discussed only monopoles in these examples, in the case of $U(N)$ gauge theories these arguments proceed identically in the case of dyons, except that dyon masses have an extra contribution coming from their electric charge, and also the above selection rules are further constrained by the fact that processes must preserve electric charge.  Similarly we may turn on a theta angle and thus shift the mass.  This is not visible in the IIA brane cartoon, but in a lift to M-theory it is simply the result of a relative displacement of the NS5-branes along the M-theory circle.

\subsection{SO and SP theories}

Traces of odd powers of elements of the $so$ or $sp$ Lie algebras are zero, and so superpotentials in $SO$ and $SP$ gauge theories must be even functions of the adjoint chiral multiplet.  In addition no FI term is possible in such theories.

Both of these facts are clear from the following brane cartoon construction.  We may make an $SO(2N)$, $SO(2N+1)$ or $SP(N)$ cartoon from the above $SU(2N)$ cartoons by adding an orientifold 4-plane parallel to the D4-branes at $x^{4,5}=0$ and $x^{7-9}=0$.  Discrete fluxes determine the boundary conditions and therefore which gauge theory we have constructed. However in every case the NS5-branes must both be symmetric under a reflection across the orientifold plane.  As the $x^7$ positions of the two branes are constant, and the $x^6$ positions are different, both $x^7$ positions must be zero and so no FI term is allowed.  We have seen that $x^8+ix^9=\W\p(x^4+ix^5)$ and so $\W\p$ must be odd to ensure that the reflection flips the signs of both $x^8+ix^9$ and $x^4+ix^5$.  This means that the superpotential, as seen in field theory, is even.  In particular
\beq
\W\p(0)=0\ ,
\eeq
and so there are no vortices along the orientifold plane.  That is to say that the squark flavors with no bare mass and the photon colors with a vanishing VEV cannot be used to construct vortices.

In the absence of an FI term the BPS confined monopoles of Refs.~\cite{HT,tong-monopolo} do not exist, although they have a BPS cousin when there are two locked flavors of hypermultiplets with distinct bare masses $m_1\neq m_2$ such that
\beq
\W\p(m_1)=\W\p(m_2)\ .
\eeq
The 2-monopole constructed using these two locked pairs is then confined by two vortices of equal tension which are mutually BPS when they are antiparallel, which is the condition that they exert no net force on the monopole.  This allows the configuration to preserve a supercharge.  If we restrict attention to the $U(2)$ theory (which may result, for example, from the breaking $SO(4)\rightarrow U(2)$) described by these two locked pairs then this monopole is related by an $SU(2)_R$ rotation to the original 1/4 BPS confined monopole of \cite{HT,tong-monopolo}.

Away from the orientifold plane everything is as in the $U(N)$ case, the only difference is that we must impose the orientifold plane's boundary conditions.  This means that there are two new types of monopoles to consider.  There may be monopoles that extend from a color brane to its reflection, and there may also be monopoles that extend from a color brane to the orientifold plane.

D2-branes may only extend from a color brane to its reflection in an $SO(2N+1)$ gauge theory.  The reflection symmetry ensures that the D4 color brane and its mirror image are both either locked or unlocked, and that if they are locked then they are both equidistant from the NS5-brane. The mirror image of a vortex occupies the same $x^{1-3}$ coordinates as the original and so does not yield a separate second vortex, therefore such monopoles are either 0-monopoles or 1-monopoles.
 D2-branes bound to a fundamental string may be suspended between a color brane and its reflection in an $SP(N)$ gauge theory, leading to $(1,1)$-dyons that behave like the monopoles in the $SO(2N+1)$ theory.


 Conversely monopoles in the $SP(N)$ theory and dyons in the $SO(2N+1)$ theory may end on the orientifold plane.  The orientifold plane supports no vortices and so such monopoles may only be 0 or 1-monopoles in this case as well.

\begin{figure}[ht]
\begin{center}
\leavevmode
\epsfxsize 9   cm
\epsffile{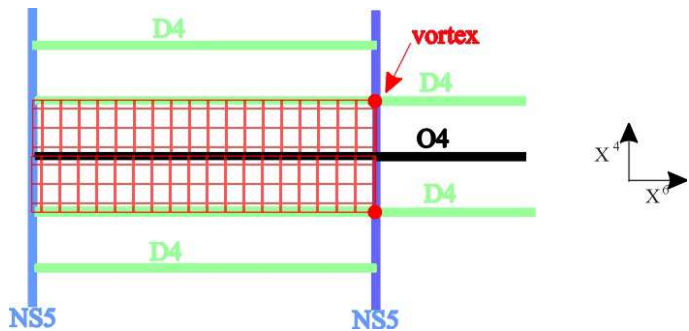}
\end{center}
\caption{\footnotesize
A 2-monopole that goes from a plane to its reflection in a $SO(2N+1)$ theory.  The orientifold projection implies that the two vortices are coincident and so these appear to be 1-monopoles.}
\label{esemp4}
\end{figure}

\subsection{What are the unconfined charges?}

For each locked color-flavor pair there is a single kind of vortex, a strip of D2-brane suspended between the NS5-brane and the locked D4-brane.  However for each such vortex there are in general many different kinds of object confined.  For example a D2-brane corresponding to a 1-monopole extends from a locked color D4-brane to an unlocked color brane.  The 1-monopole may form a boundstate with any unconfined 0-monopole that uses this unlocked color brane and a second unlocked color brane, yielding a new 1-monopole confined by the same vortex.  In all the number of kinds of 1-monopoles confined by each vortex is then equal to the number of stacks of unlocked color branes.  Semiclassically this is equal to the number of extrema of the superpotential at which there is a nontrivial gauge symmetry.  Quantum-mechanically some of these extrema may support multiple kinds of monopole as a result of dynamical symmetry breaking.  Of course in the presence of an FI term there are no unlocked colors and no 1-monopoles.

In addition to bound states with other monopoles we may also form dyons, which are bound states with W-bosons.  These are constructed by letting the boundary of the D2-brane wind around the M-theory circle along one color brane and then unwind along the other color brane.  This winding does not affect the part of the brane near the NS5-brane, and so dyons are confined by the same vortices as the monopoles that carry the same magnetic charge.

We may now construct unconfined boundstates consisting of chains of dyons connected by vortices and modify the charges by forming bound states with W-bosons and unconfined 0-monopoles.  This leads to the unsurprising result that the unconfined charges are just the unconfined monopoles plus the W-bosons.  For example in a
\beq
U(3)\longrightarrow U(1)^3\ ,
\eeq
theory in which one of the $U(1)$'s is locked we find a single kind of vortex and three kinds of monopoles, a 0-monopole $A$ and two 1-monopoles named $B$ and $C$.  Then the boundstate of a $B$ and an anti-$C$ is not confined and carries a single unit of $A$ charge.

\section{Quivers} \label{faretrasezione}


\begin{figure}[ht]
\begin{center}
\leavevmode
\epsfxsize 12  cm
\epsffile{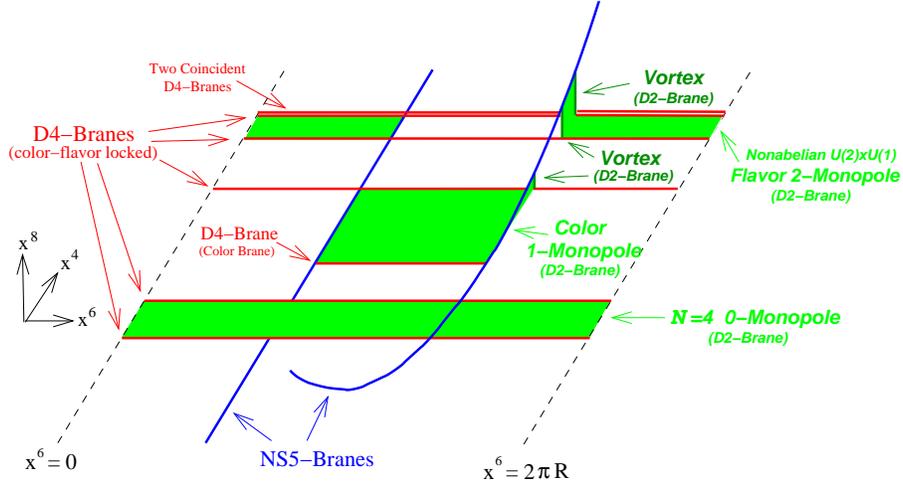}
\end{center}
\caption{\footnotesize This is a type IIA brane realization of a $U(7)\times U(6)$ quiver theory with degenerate superpotential $W\sim\Phi^3$.  A color 1-monopole,  nonabelian $U(2)\times U(1)$ flavor 2-monopole and ${\mathcal N}=4$ color-flavor 0-monopole are shown, although several other monopoles exist.  A single cross-section of the gauge theory space $x^{1-3}$ is seen here, but the three vortices each extend into one of the gauge theory directions, while the monopoles do not extend into any of the gauge theory spatial directions.}
\label{faretra}
\end{figure}

Quiver gauge theories may be obtained by compactifying the $x^6$ direction, so that it is a circle of radius $R$.  The semi-infinite flavor branes are then finite and stretch from one NS5-brane to the other, leading to the gauging of the flavor symmetry and to a $U(M)\times U(N)$ quiver theory.  The gauged flavor symmetry leads to the existence of new monopoles which are D2-branes suspended between the flavor branes.  We will refer to these as flavor monopoles to distinguish them from the old color monopoles.  If we T-dualize with respect to the 6-circle then the two NS5-branes lead to an $A_1$ singularity that is blown up by a two-sphere whose area at each point $v=x^4+ix^5$ is semiclassically
\beq
A\sim \sqrt{\W\p(v)^2+r^2}\ , \label{area}
\eeq
where $2r$ is the FI term.   Eq.~(\ref{area}) is a consequence of the fact that the sphere is semiclassically a capped cylindrical tube over the length $\sqrt{\W\p(v)^2+r^2}$ line segment between the $x^{7-9}$ coordinates of the two NS5-branes, and the tube has a constant radius $\alpha\p/R$.  The sphere also supports a $B$-field with flux
\beq
\int_{S^2} B\sim \frac{1}{2}+\frac{\Delta{x^6}(v)}{R} \ \label{b}
\eeq
where $\Delta{x^6}$ is the distance between the NS5-branes at $v$, which is roughly the inverse gauge coupling squared at the energy scale $|v|$.  Notice that both sides of Eq.~(\ref{b}) are only defined\footnote{The cascade works because the integral part of $B$ is in fact a physical quantity, it appears for example in the construction $G_5=dC_4+B\wedge G_3$ which relates it to D3-brane charge \cite{MeMonodromy}.  The torsion part of $B$ is distinguished because, while not RG invariant, it is the same when calculated in the different Seiberg-dual characterizations of the gauge theory at a fixed scale, equivalently it is invariant under the $(B+F)$-preserving large gauge transformations of the D5-brane Born-Infeld theory.  This calculation is more subtle for nonbaryonic root vacua because the effective $G_3$ is characterization-dependent \cite{MeMay}.} modulo 1 and a shift by 1 on either side corresponds to a step in the Klebanov-Strassler cascade \cite{KS}.

In the IIA brane cartoon locked color-flavor pairs are D4-branes that wrap the 6-circle, and so are T-dual to D3-branes that intersect the blown up sphere at a point and extend in the gauge theory directions.  Unlocked colors and flavors are dual to D5-branes that wrap the sphere and extend along the gauge theory directions.  These each carry a D3-brane charge that is $F+B$ cupped with the cohomology class dual to the wrapped 2-sphere, where $F$ is the fieldstrength of the D5-brane's worldvolume $U(1)$ gauge field.  The nontrivial second Stiefel-Whitney class of the normal bundle of the 2-cycle leads to a shifted quantization condition for the $U(1)$ fieldstrength
\beq
\int_{S^2} F\in \Z +\frac{1}{2} \ .
\eeq
The contribution of the $B$-field is canceled, up to an integral part, by a bulk contribution to the D3-charge \cite{Wati}.  While the $B$-field is only defined modulo an integer, the above integral contribution needs to be lifted to integral cohomology to calculate the D3-brane charge.  This lifting depends on a gauge choice, and in Ref.~\cite{MeMay} it was shown that different choices correspond to the effective theories that are weakly coupled at different energy scales, for example different choices related by large gauge transformations may be related by Seiberg duality.  This is no surprise, the number of D3-branes is the rank of the gauge group, and different choices of gauge group are weakly coupled at different energy scales.

\begin{figure}[ht]
\begin{center}
\leavevmode
\epsfxsize 12   cm
\epsffile{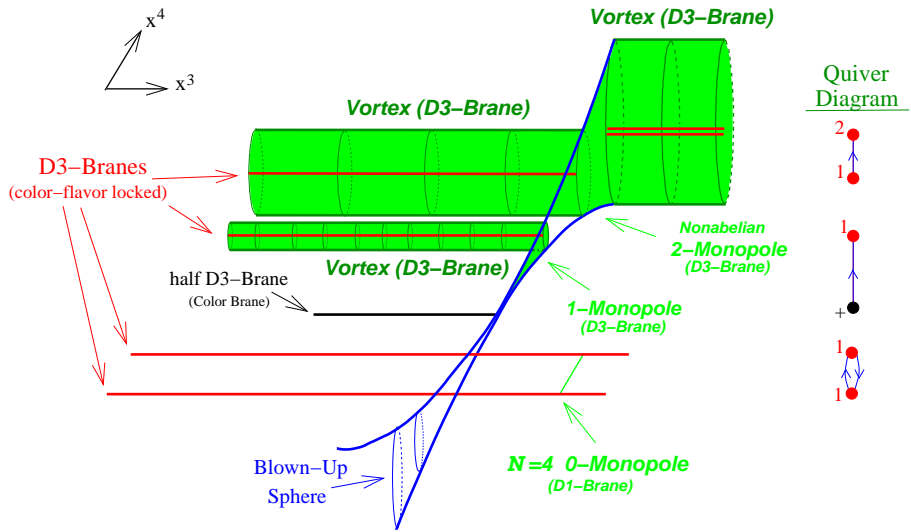}
\end{center}
\caption{\footnotesize This is a type IIB brane realization of the $U(7)\times U(6)$ quiver theory configuration seen in Fig.~\ref{faretra}.  The superpotential $W\sim\Phi^3$ now yields an $x^{4,5}$-dependent blow up of the $A_1$ singularity.  The color 1-monopole and nonabelian flavor 2-monopole are D3-branes with a half-unit of D1 charge each.  They wrap the blown up $S^2$ and so cannot be bounded by the D3 color branes, thus they continue into the $x^3$ direction as vortices.  The ${\mathcal N}=4$ color-flavor 0-monopole is a D-string.  On the right is the corresponding quiver diagram.}
\label{faretrab}
\end{figure}

To summarize, our IIB configuration is a 4-dimensional Minkowski space in which the gauge theory lives crossed with the plane $v=x^4+ix^5$ fibered over which is a resolved $A_1$ singularity whose blown-up sphere has a complex Kahler class that depends on the coupling and superpotential at $v$.  Locked color-flavor branes are D3-branes while unlocked color and flavor branes are D5-branes that wrap the sphere with a D5-brane charge (orientation) that depends on whether they were colors of flavors before compactification and a half-integral D3-charge.  A marginally locked color and flavor pair is possible where the sphere degenerates.  The pair may be separated by perturbing the superpotential.  This leaves a D5, anti-D5 pair which have no net D5-charge and a unit of D3-charge and so carries the same total charge as the locked brane to which it may be continuously deformed.

\begin{table}[h]
\begin{center}
\begin{tabular}{|l|l|l|l|}
\hline \textbf{Type of Object} & \textbf{ IIA Description} & \textbf{ IIB Description} & \textbf{Quiver Diagram} \\
\hline {Color Brane}& D4-Brane & D5 with $1/2$ D3& Black Dot Plus \\
\hline {Flavor Brane} & D4-Brane & {$\overline{\textup{D5}}$} with $1/2$ D3& Black Dot Minus\\
\hline {$n$  Locked Branes} & $n$ D4-branes & $n$ D3-branes & Red Dot with $n$ \\
\hline  &  2 NS5-Branes & $A_1$ Singularity  & \\
\hline  $W$ and FI term & NS5 Displacement & Blowup Modulus & \\
\hline  Color Monopole & D2 in $x^{4,5}$ and $x^6$ &  D3 with $1/2$ D1& Oriented Edge\\
\hline  Flavor Monopole & D2 in $x^{4,5}$ and $x^6$ &  {$\overline{\textup{D3}}$} with $1/2$ D1& Oriented Edge\\
\hline  ${\mathcal N}=4$ Monopole & D2 in $x^{4,5}$ wraps $x^6$ & D1 & Loop+2 Red Dots\\
\hline  Vortex &  D2 in $x^{7-9}$ and $x^{1-3}$& D3 on $S^2$ + $x^{1-3}$ & Edge + Red Dot\\
\hline
\end{tabular}
\end{center}
\caption{\footnotesize  Objects in IIA and IIB brane constructions and quiver diagrams }
\label{faretratavolo}
\end{table}

Monopoles are D3-branes that wrap the 2-sphere and extend between two D3 or D5 color branes.  Again fractional flux quantization on their worldvolumes leads to half-integral D1 charges.  In fact a color and flavor 2-monopole, which necessarily have opposite D3-charge, may combine to yield an unconfined 0-monopole realized by a D-string if both of their corresponding colors and flavors are locked.  These D-strings, which are 't Hooft-Polyakov monopoles of the locked color-flavor group, are stretched between two D3-branes whose positions are determined by meson VEVs.  At generic values of meson VEVs the locked D4-branes do not intersect the NS5s in IIA, thus the D3-branes and so the D-strings move away from the blown up singularity and in the infrared they are monopoles in an ${\mathcal{N}}=4$ supersymmetric subsector of the theory.

The locked D3 color branes do not wrap the sphere and so a color or flavor monopole, as it wraps the sphere, cannot actually end on such a brane unless the sphere degenerates, as in the marginally locked case.  However it may intersect such a color brane and then continue in one of the spatial directions of the gauge theory by following the brane.  This continuation is a vortex.  A vortex may follow a D3-brane for some distance and then leave it for another D3-brane which it may then follow next.  The trip between the two D3-branes carries a unit of 2-monopole charge and so this is a kink in the vortex worldsheet theory that separates two different types of vortex made using two different squarks.  The vortex tension may be calculated  to be the area of the wrapped 2-sphere, which as seen in Eq.~(\ref{area}) agrees with the vortex tension calculated above using field theory and brane cartoons, although we are now considering a different class of gauge theories.  A monopole, since it wraps the sphere, may only end on a D5 as the D5's also wrap the sphere.  That is to say that a monopole is only unconfined if it is formed from two unlocked colors.  Recalling that nonabelian gauge symmetries are dynamically broken in the absence of color-flavor locking if the superpotential is nondegenerate, we may also conclude in quiver theories as in those considered above that a color or flavor monopole transforming under $H_1\times H_2$ is confined by atleast the number of vortices equal to the number of $H_i$'s that are nonabelian.  Of course in the marginally locked case a vortex may delocalize out of existence, but the marginal locking condition receives quantum corrections and so marginal locking requires a difficult fine tuning.

We may now graphically classify monopoles in these theories.  For each stack of $n$ D3-branes we use a red vortex labeled by the number $n$ and for each D5-brane we draw a black vortex with a $+$ or $-$ depending on the orientation of the D5-brane, which depends on whether it was originally a color or a flavor.  Stacks of D5-branes will not be coincident once we allow quantum corrections by lifting to F-theory if our superpotential is nondegenerate, although more exotic superpotentials may lead to finite-coupling superconformal vacua in which there is an enhanced symmetry.  The nonexistence of stacks of D5-branes is a result of the dynamical symmetry breaking of pure Yang-Mills theories.  Notice that we cannot necessarily distinguish D3-branes from D5-branes semiclassically when the superpotential and FI-term are not much larger than the dynamical scale $\Lambda$ because the number of D3-branes is equal to the number of locked colors which depends on the vacuum chosen by the quantum theory.  For example it is equal to $r$ in the case of the nonbaryonic $r$-vacua \cite{APS}.  In the case of marginally locked pairs we must consider the D3 to be a D5, anti-D5 pair for the purpose of counting vortices, although these D5's are unusual in that they can stack.

Monopoles are oriented edges connecting the vortices.  In particular 0-monopoles connect D5-branes and so the fact that D5-branes do not appear in stacks implies that unconfined monopoles are always abelian in theories with a nondegenerate superpotential, although this observation does not extend to the marginal case.  The color-flavor bound state 0-monopoles described above, because ${\mathcal N}=4$ SUSY prevents dynamical symmetry breaking, yield a second variety of potentially nonabelian 0-monopole.  1-monopoles connect a D3 and D5-brane and so may transform in the fundamental representation of an  arbitrary unbroken component of $G$ and also a $U(1)$.  2-monopoles connect two D3-branes and so transform in the bifundamental representation of two unbroken components.  Monopoles cannot be bounded by a flavor and color brane.  This means that there are no monopoles connecting two black vortices with different signs.

A number of boundstates can be represented by paths made from their constituent edges.  The ordering of the edges, which is consistent with the edge orientation, is easily seen in the IIA realization of a given boundstate as a D2-brane disc.  The boundary of the disc is a loop that follows NS5-branes and D4-branes.  The edges of the diagram represent the segments of the path between D4-branes, while the ordering is the order in which the loop reaches each D4-brane.  In fact, if one ignores stability, bound states are classified by the fundamental group of the M5-brane as the charge of a monopole is given by the trajectory swept out by its boundary.  The simplest variety of boundstate, the color-flavor boundstates discussed above, consist of an arrow extending between two red vortices in each direction.

\section{Conclusion}


We have found the vortex tensions and number of vortices confining each monopole in $U(N)$, $SO(N)$ and $SP(N)$ gauge theories with nondegenerate superpotentials and, in the case of $U(N)$ theories, a possible FI term.  In addition we have found the conditions under which sets of vortices are mutually BPS and we have found selection rules for their decays into other vortices.  In theories with an FI term all monopoles are confined by 2 vortices, in the absence of an FI term there may also be 1-monopoles confined by a single vortex and also unconfined 0-monopoles.  However in the absence of fine-tuning we have seen that nonabelian monopoles must be confined by a number of vortices at least equal to the number (which is at most 2) of nonabelian components under which it is charged.  Futhermore we have seen that 2-monopoles are in general confined by two vortices of unequal tension and thus the vortex-monopole junction can rarely be made BPS.  For example in the absence of an FI term the imbalance of the tensions will lead to a monopole acceleration of order $g_{YM}^2\mu$ even when the vortices pull in opposite directions, where $\mu$ is the adjoint chiral multiplet mass when expanded about the eigenvalues used in the construction of the monopole (which are taken to be close to each other in this approximation).

In the case of quiver theories, in which the flavor symmetry is gauged, the situation is different.  In addition to monopoles formed from the color symmetry, there are also monopoles formed from the gauged flavor symmetry.  If the monopoles are made from colors that are all locked to flavors, a color and flavor 2-monopole may form a boundstate that is unconfined and nonabelian and in general described by an effective ${\mathcal N}=4$ subsector the theory.  In a type IIA and IIB description of these theories we see that the vortex tension formula is the same as the vortex tension formula for theories in which flavor is not gauged.

The MQCD approach used on $U(N),\ SO(N),\ SP(N)$ and quiver theories above appears not to apply to $SU(N)$ gauge theories.  For example the 't Hooft-Polyakov monopoles of $SU(2)\rightarrow U(1)$ gauge theories transform under a single $U(1)$, and so there is only one type of vortex in such theories even when both colors are locked.  Intuitively we might construct such a vortex by taking the 2-monopole of a $U(2)\rightarrow U(1)^2$ theory and quotienting by a diagonal $U(1)$, and so the single vortex of the $SU(2)$ theory is roughly a linear combination of the two vortices of the $U(2)$ theory.  However we know that such vortices are not always mutually BPS, and so unlike the above cases it is possible that $SU(2)$ theories often do not admit BPS vortices.  We will begin an investigation of these vortices and connect with the solutions of Ref.~\cite{vortici} in \ref{suapp}.

\appendix

\section{Nonabelian Vortex} \label{NAV}

Now we study the nonabelian BPS vortex and we see that it is classically a simple generalization of the U(1) case. We consider $N_c$ colors and $N_f \geq N_c$ flavours.
The relevant terms in the Lagrangian are:
\beq
{\cal L}\supset -\frac{1}{2g^2}\Tr_{N_c} \, (F_{\mu\nu}F^{\mu\nu}) - (D_\mu q)(D^\mu q)-\frac{g^2}{4}\Tr_{N_c}\,((qq\dag -2{r}{\bf 1}_{N_c})^2)\ .
\eeq
The potential is minimized when
\beq
<q>=\left(\begin{array}{ccccc}
\sqrt{2{r}}&&&0&\\
&\ddots&&&\ddots\\
&&\sqrt{2{r}}&&\\
\end{array}\right) \ .
\eeq
This VEV breaks the global symmetries of the Lagrangian
\beq
\label{ressimm}
U(N_c) \times U(N_f) \rightarrow U(N_c)_{C+F} \times U(N_f-N_c) \ .
\eeq

We write the tension of the vortex using the Bogomolny trick
\beq
T=T_{BPS}+ \int d^2x  \, \frac12 (D_k q+i\epsilon_{kl}D_l q)\dag (D_k+i\epsilon_{kl}D_l q)
+\frac12 \Tr_{N_j}\, (\frac 1g F_{kl}+\frac{g}{2}(qq\dag -2{r}{\bf 1})\epsilon_{kl})^2 \ ,
\eeq
where the lower bound is
\beq
T_{BPS}=-\int d^2x \, \epsilon_{kl}\Tr_{N_j}\,(i(D_lq)( D_k q)\dag +\frac12 F_{kl}(qq\dag -2{r}{\bf 1}))\ .
\eeq
After some manipulation we can show that $T_{BPS}$ is a boundary term
\beq
T_{BPS}=\int d^2 x \, \epsilon_{kl} \partial_k \Tr_{N_j}\,(2{r} A_l-i(D_lq)q\dag )\ .
\eeq
To have a finite energy configuration $D_l q$ must go rapidly to zero, so the BPS bound is
\beq
T_{BPS}=2{r} \oint d \vec x \cdot \Tr_{N_j}\, \vec A \ .
\eeq
This term saturates the tension when the nonabelian Bogomolny equations
\beq
D_kq+i\epsilon_{kl}D_l q=0 \ ,\qquad \frac 1g F_{kl}+\frac{g}{2}(qq\dag -2{r}{\bf 1})\epsilon_{kl}=0
\eeq
are satisfied.  Note that the first is an equation for each squark flavor and the second  is a matrix relation.

To build the vortex configuration we embed the ordinary $U(1)$ vortex in this theory.  All such embeddings are $U(N_c)$ rotations of
\bea
\label{vortexconf}
q&=&\left(\begin{array}{cccccc}
e^{i\theta}\sqrt{2{r}}q(r)&&&&0&\\
&\sqrt{2{r}}&&&&\ddots\\
&&\ddots&&&\\
&&&\sqrt{2{r}}&&\\
\end{array}\right)\ ,\\
A_k&=&\left(\begin{array}{cccc}
-\epsilon_{kl}\frac{\hat{r}_l}{r}f(r)&&&\\
&0&&\\
&&\ddots&\\
&&&0\\
\end{array}\right) \ ,\nonumber
\eea
where $q(r)$ and $f(r)$ are some profile functions that satisfy the boundary conditions $q(0)=f(0)=0$ and $q(\infty)=f(\infty)=1$.
The $N_c$ independent vortices constructed this way are degenerate with tension
\beq
T_{BPS}=4\pi{r} \ .
\eeq

The vortex solution (\ref{vortexconf}) classically breaks the residual global symmetry (\ref{ressimm}). This leads to the existence of a moduli space. When $N_f=N_c$ the moduli space is compact:
\beq
\mathbb{CP}^{N_c-1}=\frac{ U(N_c)_{C+F} }{ U(1) \times U(N_c-1)} \ .
\eeq
When $N_f>N_c$ other zero modes are present in our theory.
Hanany and Tong
\cite{HT,HT2} have found that among the classical solutions for $N_f>N_c$ are semi-local vortices, which were first discovered
in abelian gauge theories with multiple Higgs fields \cite{semilocal1,AV}.
These solitons interpolate between Nielsen-Olesen like vortices
and sigma-model lumps on the Higgs branch of the theory.
The moduli space of semilocal vortices is not compact;
given $N_c$ colors and $N_f$ flavor in the low energy
effective action, the moduli space is
the following manifold \cite{HT2}:
\beq \sum_{i=1}^{N_c}|\phi_i^2|-\sum_{j=1}^{N_f-N_c}|\tilde{\phi}_j^2|=1,
\eeq
divided by a $U(1)$ action in which the $\phi_i$
fields have charge $1$ and the $\tilde{\phi}_j$ have charge $-1$.
This moduli space reduces to $ {\mathbb{CP}}^{N_c-1}$ for $N_f=N_c$.

\section{Vortex in the $\mathbf{SU(N_c)}$ Theory } \label{suapp}

We now turn to $SU(N_c)$ theories, again with
nondegenerate superpotential $W(\Phi)$ (there is no Fayet-Iliopoulos  term
as the gauge group has no abelian factors) and $r$ locked colors.
The VEVs can be computed by
 adding a Lagrange multiplier:
\beq
\widetilde{\W}=\W+\nu \, \Tr_{N_c} \, \Phi \ .
\eeq
Minimization with respect to $\nu$ leads to the condition $\Tr_{N_c} \, \Phi = 0$.
We impose
$\widetilde{W}'(\phi_i)=0$ for all of the unlocked colors $\phi_{r+1}
\, \ldots \,\phi_{N_c}$.
Then a system of algebraic equations is found
\beq \nu= -W'(\phi_{r+1})= \ldots = -W'(\phi_{N_c})\ , \qquad
\sum_{k=r+1}^{N_c} \phi_k= - \sum_{l=1}^{r} m_l  \ . \eeq
Note that we can find many solutions in which the adjoint scalar has degenerate eigenvalues.
These solutions are modified by quantum mechanical
effects: there is dynamical symmetry breaking
because there are no massless quarks hypermultiplets that make
the nonabelian effective theory infrared free.
In the case $r=N_c$
the Lagrange multiplier $\nu$ is undetermined.
The meson VEVs are:
\bea
& \widetilde{q}_i q_i = \widetilde{W}'(m_i)\ , \qquad & 1\leq i \leq r\ , \\
&  \widetilde{q}_i q_i=0 \ , \qquad &  i > r \ . \nonumber
\eea

Let us choose an orthogonal basis $T^j$
of $SU(N_c)$'s Cartan subalgebra (acting as a matrix on $\vec{v},\vec{w}$)
with $\Tr_{N_c}\, T^j T^i= \delta_{ij} /2$.
The basis is chosen such that
there are no mixed $U(1)$ gauge couplings $\tau_{ij}$
for $i \neq j$.
The effective bosonic Lagrangian describes
a $U(1)^{N_c-1}$ theory with $r$ scalars $q_i$, $\widetilde{q}_i$:
\beq \sum_j -\frac{1}{4 e_j^2} |F^j_{\mu \nu}|^2
- \sum_i (|D_\mu q_i|^2 + |D_\mu \widetilde{q}_i|^2) - V_D -V_F \ ,
\label{metalmeccanico} \eeq
where
\bea
&  V_D=\sum_j \frac{e_j^2}{2} | T^j \vec{w} |^2\ , \qquad
 & \vec{w}_k=|q_k|^2-|\widetilde{q}_k|^2 \ , \\
& V_F=  \sum_j 2 e_j^2 | T^j \vec{v}|^2 \ , \qquad
&  \vec{v}_k=(\widetilde{q}_k q_k-\widetilde{W}'(m_k))\ . \nonumber
\eea

Using this Lagrangian the $U(1)^{N_c-r-1}$ subgroup
of gauge bosons constructed from only the unlocked colors
decouples from the other low energy bosonic fields.
In particular, the vortex soliton will not contain these fields.
The effective theory in which we look for vortex solutions is
then a $U(1)^{N_c-1}$ gauge theory for $r=N_c$ and
a  $U(1)^{r}$ for $1 \leq r \leq N_c-1$, with possible nonabelian symmetry for the locked colors as usual.

\subsection{$\mathbf{SU(2)}$ case}

Let's consider the simplest case $N_c=2$. When $\widetilde{\W}'(\phi_1)=\widetilde{\W}'(\phi_2)=0$
there are no vortices and the monopole is unconfined. When only one color is locked $\phi_1=m$ and $\widetilde{\W}'(-m)=0$, the relevant terms are
(we use the ansatz $\widetilde{q}_1=q_1^\dagger e^{i \theta} $ where
$\theta$ is the phase of $A=-\widetilde{W}'(m)=-(W'(m)-W'(-m))$):
\beq
 -\frac{1}{4e^2}F_{\mu\nu}F^{\mu\nu} -
 \sum_{i=1}^2 2(|D_\mu q_1|^2)
 -\frac{e^2}{2}( q_1^\dagger q_1-|A|)^2 \ .
\eeq
We find an ordinary $U(1)$ vortex of charge $2\pi\left(\begin{array}{cc} 1&\\&-1\\ \end{array}\right)$
that, unlike the case of $U(N)$ 1-monopoles, confines all of the flux of the monopole. The tension of this vortex is
\beq T_{BPS} = 4\pi|\widetilde{\W}'(m)|\ . \eeq

The two colors may be locked if there are two flavours with bare masses $m$ and $-m$.  The low energy theory has a $U(1)$ gauge symmetry and two charged hypermultiplets.
The following ansatz can be used:
\beq \widetilde{q}_1 = q_1^\dagger e^{i \theta}, \,
q_2 = -\widetilde{q}_2^\dagger e^{i \theta} \ ,\eeq
where $\theta$ is the phase of $A=-(W'(m)-W'(-m))$.
The relevant part of the Lagrangian is:
\beq
 -\frac{1}{4e^2}F_{\mu\nu}F^{\mu\nu} -
 2(|D_\mu q_1|^2)-2(|D_\mu \widetilde{q}_2|^2)
 -\frac{e^2}{2} (q_1^\dagger q_1 +\widetilde{q}_2^\dagger \widetilde{q}_2-|A|)^2\ .
\eeq
The fields $q_1$, $\widetilde{q}_2$ have the same $U(1)$
charge and there is a global $SU(2)$ symmetry
which rotates $(q_1,\widetilde{q}_2)$ as a doublet.
We can use this global symmetry
to impose the singular gauge $(q_1,\widetilde{q}_2)\cong(0,|A|)$
at infinity.
This model was already analyzed for
example in \cite{AV}: semilocal strings
are obtained.
A continuous family of solutions
can be found in our BPS case \cite{AV}.
For the vortex with winding number $1$
the explicit solution is
(polar coordinates $(r,\phi)$ are used)
\beq \left( \begin{array} {c} q_1  \\ \widetilde{q}_2
\end{array} \right) = \frac{1}{\sqrt{r^2+|v|^2}}
 \left( \begin{array} {c} v  \\ |A| r e^{i \phi}
\end{array} \right) e^{ u(r,|v|)/2 },
\eeq
where $u$ is the solution to
\beq \nabla^2 u + 2(1-e^u)=\nabla^2 \ln (r^2+|v|^2)\ ,\eeq
with $u \rightarrow 0$ as $r \rightarrow \infty $.
The complex parameter $v$ is a coordinate in the vortex moduli
space; for $v \neq 0$ the magnetic field tends to zero
as $B \approx 2 |v|^2 r^{-4}$.
The tension of this family of strings is:
\beq T_{BPS}=4 \pi |A|=4 \pi |(W'(m)-W'(-m))|\ .\eeq
The flux of these vortices is the same as the monopole flux,
so the monopole is confined by one (semi-local) vortex.

\subsection{Some remarks on the $SU(N_c)$ case}

The effective theory in which we look for vortex solutions is
a $U(1)^{N_c-1}$ gauge theory for $r=N_c$ and
$U(1)^{r}$ for $1 \leq r \leq N_c-1$.
In general the low energy effective
couplings are distinct (see \ref{metalmeccanico}) and so
they are diagonal only in a particular basis $T^n$
of the $SU(N_c)$ Cartan algebra.  In the $SU(N_c)$ theory only traceless states exist,
so  the fundamental vortices of the $U(N_c)$ are no longer present.
As a result a vortex needs to carry charge in at least two colors.

When the complex phases of $\widetilde{W}'(m_k)$ are different,
it is not clear how to generalize
the previous $SU(2)$ ansatz
 $\widetilde{q}=q^\dagger e^{i \theta}$;
perhaps in general
 such vortices are not BPS as they resemble quotients of superpositions of non mutually BPS vortices in a $U(N)$ subsector.
On the other hand, if the complex
phases of $\widetilde{W}'(m_k)$ are all the same and $r<N_c$
it is easy to generalize the approach used for $SU(2)$.
We can first take all of the $\widetilde{W}'(m_k)$ to be real and positive
using the appropriate $SU(2)_R$ rotation;
then the following ansatz is used:
\beq  \widetilde{q}_k=q_k^\dagger
\eeq
for all of the squarks fields $\widetilde{q}_k$.
The squark profile for the $l$th vortex is
\beq q_l=e^{i \varphi} \phi_l (r) \ ; \qquad
 q_k= \phi_k (r) \ , \,\, k \neq l\eeq
where $\phi_k(r)$ are real functions such that $\phi_k(r\rightarrow\infty)$ is $\sqrt{\widetilde{W}'(m_k)}$ when $k\leq r$ and otherwise vanishes.
The vortex charge is (where 1 is
the $l$-th entry and $J=N_c-r$)
\beq
Q^{(l)}=2 \pi \left(\begin{array}{ccc|c}
\ddots&&&\\
&1&&\\
&&\ddots& \\
\hline
&&&-\frac{1}{J} \bf{1}_J\\
\end{array}\right) \ .
\eeq

This ansatz can be used in the Hamiltonian
derived from the Lagrangian (\ref{metalmeccanico}).
A convenient redefinition of the squark fields
$q_k \rightarrow \frac{1}{\sqrt{2}} q_k$ is used
in the following equations, here
$i$ and $j$ are space indices and the gauge fields
are written in the $T^n$ basis ($F_{12}=F_{12}^{(n)} \, T^n$).  We first introduce
 \beq s_k= q_k^\dagger q_k
-2 \widetilde{W}'(m_k) \ , \qquad u_k=2 \widetilde{W}'(m_k) \ . \eeq
Now the string tension can be written \`a la Bogomolny \cite{Bogomolny}:
\beq  T= \int dx^2 \,\left( \sum_n \left[ \frac{1}{2 e_n} F^{(n)}_{ij}+
\frac{e_n}{2}  ( \, \Tr \, (T^n \vec{s} ) )
\right]^2  +\frac{1}{2}|\nabla_i q_k + i \epsilon_{ij} \nabla_j q_k|^2 \right)
\eeq
\[  +\int dx^2 \,
\left ( \sum_n \, \Tr \, (T^n \vec{u}) \,
 2 \left[ \frac{1}{2} \epsilon_{ij} F^{(n)}_{ij} \right]  \right) \
\]
and the BPS equations are
\beq   F^{(n)}_{12} + e_n^2 ( \, \Tr \, (T^n \vec{s} ) )=0 \ ,
\eeq
\beq \nabla_i q_k + i \epsilon_{ij} \nabla_j q_k =0 \ . \eeq
The tension of this vortex is given by
the topological term
\beq T^l_{BPS}=4 \pi  | \sum_n  \, \Tr \, (T^n \vec{u}) \, \Tr \, (T^n Q^{(l)}) |
=4 \pi |\widetilde{W}'(m_l)| \ .  \label{megera} \eeq

This system has been analyzed in Ref.~\cite{MY}
for the $r=1$\ and $2$ vacua of the
 $SU(3)$ theory  with $W'(m_i)$ real and positive;
the general bound state $(n,k)$ with charge $Q = n Q^{(1)} + k Q^{(2)}$ was also
 considered. The result is that a BPS configuration
can be found only if $(n,k)$ are both positive or both negative.

When only
one color is not locked ($r=N_c-1$) it is possible to write a simple
formula for the Lagrange multiplier $\nu$.
Suppose that $\phi_i=m_i$ for $i=1\ ...\ N_c-1$;
the $\Tr_{N_c}\, \Phi =0 $ constraint gives
$ \phi_{N_c}=-\sum_{i=1}^{N_c-1} m_i$.
The condition that this last color is not
locked is $\widetilde{W}'(\phi_{N_c})=0$
where $\widetilde{W}$ is the sum of the superpotential
and the Lagrange multiplier $\nu \,  \Phi$.
This gives the following expression for $\nu$:
\beq \nu=-W'(-\sum_{i=1}^{N_c-1} m_i)\ . \eeq
This may be compared with the tension of an SU(2) charged vortex studied in
 \cite{vortici}.  The parameters $\mu$ and $m$
 here are $-\frac{1}{\sqrt{2}}$ times the ones
 in that paper.  In the conventions of that article we find
 $\widetilde{W}'(m)=3 \mu m$ and,
 using Eq.~(\ref{megera}), $T_{BPS}=12 \pi \mu m$.
This is indeed the result found in Eq.~(3.12) of \cite{vortici}.

\section* {Acknowledgement}

We would like to thank Ken Konishi, David Tong, Mikhail Shifman and Alexei Yung for illuminating discussions.

The work of JE is partially supported by IISN - Belgium (convention 4.4505.86), by the ``Interuniversity Attraction Poles Program -- Belgian Science Policy'' and by the European Commission RTN program HPRN-CT-00131, in which he is associated to K. U. Leuven.  He would also like to thank the University of Adelaide for hospitality and many delicious kangaroo while part of this work was in progress.

\end{document}